\documentclass[a4paper,11pt]{article}
\pdfoutput=1 

\usepackage{jheppub} 
\usepackage{etoolbox}
    \makeatletter
    \patchcmd{\maketitle}{\@fpheader}{}{}{}
    \makeatother                     

\usepackage[T1]{fontenc} 
\usepackage{slashed}

\newcommand{\be}{\begin{equation}}
\newcommand{\ee}{\end{equation}}
\newcommand{\bea}{\begin{eqnarray}}
\newcommand{\eea}{\end{eqnarray}}

\title{\boldmath Lightening Gravity-Mediated Dark Matter}

\preprint{CERN-TH-2019-229}

\author[]{Yoo-Jin Kang}
\author[]{and Hyun Min Lee}

\affiliation[]{Department of Physics, Chung-Ang University, Seoul 06974, Korea}
\affiliation[]{CERN, Theory department, 1211 Geneva 23, Switzerland}

\emailAdd{yoojinkang91@gmail.com}
\emailAdd{hminlee@cau.ac.kr }

\abstract{We revisit the scenario of a massive spin-2 particle as the mediator for communicating between dark matter of arbitrary spin
and the Standard Model. Taking the general couplings of the spin-2 particle in the effective theory, we discuss the thermal production mechanisms for dark matter with various channels and the dark matter self-scattering. For WIMP and light dark matter cases, we impose the relic density condition and various experimental constraints from direct and indirect detections, precision measurements as well as collider experiments.  We show that it is important to include the annihilation of dark matter into a pair of spin-2 particles in both allowed and forbidden regimes, thus opening up the consistent parameter space for dark matter. 
The benchmark models of the spin-2 mediator are presented in the context of the warped extra dimension and compared to the simplified models.
}

\begin{document} 
\maketitle
\flushbottom

\section{Introduction}
\label{sec:intro}

Dark matter (DM) is a complete mystery in particle physics and cosmology, although its presence can be unambiguously inferred from galaxy rotation curves, gravitational lensing, Cosmic Microwave Background as well as large-scale structures, etc.
There are null results in searching dark matter beyond gravitational interactions from various direct and indirect detection experiments, thus, in particular, a lot of parameter space for Weakly Interacting Massive Particles (WIMPs) has been ruled out \cite{xenon1t,lux,pandax}.  

The nature of dark matter is still an open question.
To this, it is very important to pin down the production mechanisms for dark matter in the early universe. For instance, WIMP dark matter relies on the freeze-out process under which the DM relic density is determined in terms of weak interaction and weak-scale DM mass. Thus, this has motivated specific target materials and technologies in the direct searches for WIMP for more than three decades.  New production mechanisms such as for Feebly Interacting Massive Particles (FIMPs) \cite{FIMP}, Strongly Interacting Massive Particles (SIMPs) \cite{SIMP0,SIMP} and forbidden dark matter \cite{FDM0,FDM}, etc, can motivate different target materials and new technologies to get access to sub-GeV DM masses and/or feeble interactions. It is known that light dark matter with sub-GeV mass can have large self-interactions to solve potentially small-scale problems at galaxies \cite{smallscale,smallscale2} and it may also call for new dynamics in the dark sector \cite{DMdyn} to get the DM self-interactions velocity-dependent for galaxy clusters such as Bullet cluster \cite{bullet}.

Moreover, dark matter is known to be neutral under electromagnetism, so it is conceivable to communicate between dark matter and the Standard Model (SM) through messenger or mediator particles. Thus, the simplified models for dark matter with mediator particles have drawn a lot of attention, providing an important guideline for direct and indirect detections of dark matter as well as collider experiments \cite{DM-forum0,DM-forum}.

In this article, we consider a massive spin-2 particle as the mediator for dark matter of arbitrary spin, which couples to the SM particles and dark matter through the energy-momentum tensor, as originally proposed by one of us and collaborators \cite{GMDM}.  This scenario has been dubbed ``Gravity-mediated dark matter'', due to the similarity to the way that the massless graviton interacts with the SM. The spin-2 mediator stems from a composite state in conformal field theories or a Kaluza-Klein(KK) graviton in a gravity dual with the warped extra dimension \cite{GMDM,GMDM2,GMDM3,GMDM-dd}. There are other works on the spin-2 mediated dark matter in similar frameworks \cite{GMDM-others,GMDM-other2}.
We regard the massive spin-2 particle as a dark matter mediator in the effective theory with general couplings to the SM and dark matter and discuss the general production mechanisms for WIMP dark matter and light dark matter in this scenario.

We discuss various channels of dark matter interactions in the presence of the spin-2 mediator: direct $2\rightarrow 2$ annihilations, $2\rightarrow 2$ allowed and forbidden channels into a pair of spin-2 mediators, $3\rightarrow 2$ assisted annihilations as well as DM self-scattering.  
We perform a comprehensive check of the consistency between the correct relic density and various experimental constraints, such as direct detection, precision measurement of muon $g-2$, meson decays and collider experiments, in both WIMP and light dark matter cases.
We also introduce two benchmark models with the warped extra dimension for the spin-2 mediator, such as the Randall-Sundrum(RS) model \cite{rs} and the clockwork model \cite{clockwork0,clockwork}. 
Then, we discuss the impacts of heavier KK gravitons on the aforementioned processes for dark matter, focusing on the DM $s$-channel annihilation into the SM particles and DM elastic scattering processes. 

There is a recent work \cite{G2} where a similar setup is studied for the massive spin-2 particle playing a role as a mediator for dark matter and the parameter space for heavy dark matter beyond TeV scale is scanned over in the context of 5D linear dilaton background, based on standard WIMP $2\rightarrow 2$ annihilation channels. On the other hand, in our work, we focus on the phenomenological study of the massive spin-2 mediator in the effective theory, focusing on the productions and constraints of weak-scale WIMP and sub-GeV dark matter with new production channels, and deal with the complete analysis of DM direct detection constraints.

The paper is organized as follows. 
We begin with a brief description of our setup for the spin-2 mediator and its interactions.
Then, we determine the DM relic density from various annihilation channels and discuss the self-scattering process for dark matter and the unitarity bounds. Next, we consider the DM-nucleon elastic scattering for WIMP and the DM-electron elastic scattering for light dark matter and provide various direct and indirect constraints on those dark matter models.
We continue to show two benchmark models with the warped extra dimension and discuss how the DM processes can be modified due to extra resonances. 
Finally, conclusions are drawn.
There are three appendices dealing with the details on DM-nucleon scattering amplitudes, decay widths of spin-2 particles as well as the KK sums.

\section{The setup}

We consider the effective interactions of a massive spin-2 field, $G_{\mu\nu}$, to the SM particles as well as dark matter with arbitrary spin, in the following \cite{GMDM},
\bea
{\cal L}_{\rm eff} &=& \frac{c_1}{\Lambda} G^{\mu\nu} \Big( \frac{1}{4} \eta_{\mu\nu} B_{\lambda\rho} B^{\lambda\rho}+B_{\mu\lambda} B^\lambda\,_{\nu} \Big)+\frac{ c_2}{\Lambda} G^{\mu\nu}  \Big( \frac{1}{4} \eta_{\mu\nu} W_{\lambda\rho} W^{\lambda\rho}+W_{\mu\lambda} W^\lambda\,_{\nu}  \Big)  \nonumber \\
&&+ \frac{c_3}{\Lambda} G^{\mu\nu}  \Big( \frac{1}{4} \eta_{\mu\nu} g_{\lambda\rho} g^{\lambda\rho}+g_{\mu\lambda} g^\lambda\,_{\nu} \Big) -\frac{ic_\psi}{2\Lambda}G^{\mu\nu}\left(\bar{\psi}\gamma_{\mu}\overleftrightarrow{D}_{\nu}\psi-\eta_{\mu\nu}\bar{\psi}\gamma_{\rho}\overleftrightarrow{D}^{\rho}\psi\right) \nonumber \\
&&+\frac{c_H}{\Lambda}G^{\mu\nu}\left(2(D_{\mu}H)^{\dagger}D_{\nu}H-\eta_{\mu\nu}\left((D_{\rho}H)^{\dagger}D^{\rho}H-V(H)\right)\right) +\frac{c_{\rm DM}}{\Lambda}\, G^{\mu\nu}  T^{\rm DM}_{\mu\nu}
\eea
where $B_{\mu\nu}, W_{\mu\nu}, g_{\mu\nu}$ are the strength tensors for $U(1)_Y, SU(2)_L, SU(3)_C$ gauge fields, respectively, $\psi$ is the SM fermion, $H$ is the Higgs doublet, and $\Lambda$ is the dimensionful parameter for spin-2 interactions.
Here, we note that $c_{i}(i=1,2,3)$, $c_\psi$, and $c_H$ are dimensionless couplings for the KK graviton. Depending on the spin of dark matter, $s=0,\frac{1}{2}, 1$, denoted as $S$, $\chi$ and $X$, the energy-momentum tensor for dark matter, $T^{\rm DM}_{\mu\nu}$, is given, respectively, by
\bea
T^{ S}_{\mu\nu}&=&c_S\bigg[ \partial_\mu S \partial_\nu S-\frac{1}{2}g_{\mu\nu}\partial^\rho S \partial_\rho S+\frac{1}{2}g_{\mu\nu}  m^2_S S^2\bigg],\\
T^{\chi}_{\mu\nu}&=& c_\chi \bigg[\frac{i}{4}{\bar\chi}(\gamma_\mu\partial_\nu+\gamma_\nu\partial_\mu)\chi-\frac{i}{4} (\partial_\mu{\bar\chi}\gamma_\nu+\partial_\nu{\bar\chi}\gamma_\mu)\chi-g_{\mu\nu}(i {\bar\chi}\gamma^\mu\partial_\mu\chi- m_\chi {\bar\chi}\chi) \bigg]
\nonumber \\
&&+\frac{i}{2}g_{\mu\nu}\partial^\rho({\bar\chi}\gamma_\rho\chi)\bigg],  \nonumber \\
T^{X}_{\mu\nu}&=&
c_X\bigg[ \frac{1}{4}g_{\mu\nu} X^{\lambda\rho} X_{\lambda\rho}+X_{\mu\lambda}X^\lambda\,_{\nu}+m^2_X\Big(X_{\mu} X_{\nu}-\frac{1}{2}g_{\mu\nu} X^\lambda  X_{\lambda}\Big)\bigg].
\eea

In the later discussion, we focus on the couplings of the spin-2 mediator to quarks, leptons and massless gauge bosons in the SM, as well as dark matter couplings. We treat those SM to mediator couplings to be independent parameters, but be universal for simplicity as well as unitarity consideration.

\section{Dark matter annihilations and self-scattering}

In this section, we discuss the Boltzmann equations for determining the relic density of dark matter and show the details for the cross sections for $2\rightarrow 2$ direct annihilations.
In particular, we obtain for the first time the new results for $2\rightarrow 2$ forbidden channels, $3\rightarrow 2$ assisted annihilations, and DM self-scattering.

First, we consider the Boltzmann equations for the relic density of real scalar dark matter $S$ or vector dark matter $X$, given by
\begin{equation}\begin{aligned}
\dot{n}_{\rm DM} + 3Hn_{\rm DM} =\ &-2\langle \sigma v \rangle_{{\rm DM\, DM}\rightarrow {\rm SM\,SM}}\bigg(n_{\rm DM}^2-(n_{\rm DM}^{\rm eq})^2 \bigg)  
\\
& -2\langle\sigma v^2\rangle_{{\rm DM\,DM\,DM}\rightarrow {\rm DM}\,G}\bigg(n_{\rm DM}^3-(n_{\rm DM}^\text{eq})^2 n_{\rm DM}\bigg)\\
& -2\langle \sigma v \rangle_{{\rm DM\,DM}  \rightarrow GG}\bigg(n_{\rm DM}^2- (n^{\rm eq}_{\rm DM})^2 \bigg).  \label{Boltzmann-S}
\end{aligned}\end{equation}

Similarly, for Dirac fermion dark matter $\chi$, the corresponding Boltzmann equation for $n_{\rm DM}=n_\chi+n_{\bar\chi}$ is
\begin{equation}\begin{aligned}
\dot{n}_{\rm DM}+ 3Hn_{\rm DM} =\ &-\frac{1}{2}\langle \sigma v \rangle_{\chi{\bar\chi}\rightarrow {\rm SM\,SM}}\bigg(n_{\rm DM}^2-(n_{\rm DM}^{\rm eq})^2 \bigg)  
\\
& -\frac{1}{2}\langle\sigma v^2\rangle_{\chi{\bar\chi}\chi\rightarrow \chi G}\bigg(n_{\rm DM}^3-(n_{\rm DM}^\text{eq})^2 n_{\rm DM}\bigg)\\
& -\frac{1}{2}\langle \sigma v \rangle_{\chi{\bar\chi}\rightarrow GG}\bigg(n_{\rm DM}^2-(n_{\rm DM}^\text{eq})^2 \bigg).  \label{Boltzmann-F}
\end{aligned}\end{equation}

Henceforth, we assume that the spin-2 particle is in thermal equilibrium with the SM plasma during the freeze-out, so we can take $n_G=n^{\rm eq}_G$, which is the number density in thermal equilibrium.

\subsection{Direct annihilations}

We focus on the cases with relatively light WIMP dark matter and light dark matter below the WW threshold, which annihilate dominantly into the SM fermions or massless gauge bosons. 

If dark matter is heavier than the $WW$ threshold, we can also take into account the DM annihilations into the electroweak sector, as shown in Ref.~\cite{GMDM,GMDM2}, allowing for smaller couplings of the spin-2 mediator to the SM particles for a correct relic density.  
In this work, however, for WIMP dark matter, we take the spin-2 mediator couplings to the SM quarks and gluons to be nonzero in simplified models. For consistency of gauge-invariant couplings, we choose $c_1=c_2=c_H=0$ in the electroweak sector and $c_l=0$ for SM leptons in the discussion for WIMP.  
On the other hand, for light dark matter below the $WW$ threshold, we keep all the spin-2 mediator couplings to the SM to be nonzero.

In the case when dark matter is heavier than the spin-2 mediator, dark matter can also annihilate directly into  a pair of spin-2 particles, reducing the dark matter abundance further together with the direct annihilations into the SM. 

In the case where $2\rightarrow 2$ annihilation channels are dominant,  the Boltzmann equations, (\ref{Boltzmann-S}) or (\ref{Boltzmann-F}), become
\bea
\dot{n}_\text{DM} + 3Hn_\text{DM}\approx -\langle \sigma v\rangle_{2\rightarrow 2}\, n^2_{\rm DM} \label{Boltzmann1}
\eea
with
\bea
(\sigma v)_{2\rightarrow 2}&\equiv& \left\{ \begin{array}{cc} 2( \sigma v )_{{\rm DM\,DM}\rightarrow {\rm SM\,SM}}
+2(\sigma v)_{{\rm DM\,DM}\rightarrow GG},\quad {\rm DM}=S, X, \\  \frac{1}{2} ( \sigma v)_{\chi{\bar\chi}\rightarrow {\rm SM\,SM}} + \frac{1}{2}( \sigma v)_{\chi{\bar\chi}\rightarrow GG},\quad {\rm DM}=\chi,
\end{array} \right.
 \nonumber \\
&\equiv & a+ b\, v^2+ c\, v^4.
\eea
Then, the relic density for WIMP dark matter is given by
\bea
\Omega_{\rm DM} h^2 = 5.20\times 10^{-10}\,{\rm GeV}^{-2} \bigg(\frac{10.75}{g_*} \bigg)^{1/2} \Big(\frac{x_f}{20} \Big) \bigg(a+ \frac{3b}{x_f}+\frac{20c}{x^2_f} \bigg)^{-1}
\eea
with $x_f=m_{\rm DM}/T_f$ where $T_f$ is the freeze-out temperature.

\subsubsection{Scalar dark matter}

The annihilation cross section for scalar dark matter into a pair of SM fermions, $SS\rightarrow \psi{\bar\psi}$, is given \cite{GMDM,GMDM2,GMDM3} by 
\bea
(\sigma v)_{SS\rightarrow\psi{\bar\psi} } = v^4 \cdot  \frac{ N_c c_S^2 c_\psi^2 }{360\pi \Lambda^4} 
\frac{m_S^6}{(m_G^2-4 m_S^2)^2+\Gamma_G^2 m_G^2}
\left(1-\frac{m_\psi^2}{m_S^2}\right)^\frac{3}{2} \left(3+\frac{2m_\psi^2}{m_S^2}\right)
\eea
where $N_c$ is the number of colors for the SM fermion $\psi$, and $\Gamma_G$ is the width of the spin-2 particle.
Thus, the annihilation of scalar dark matter into the SM fermions becomes $d$-wave suppressed, so scalar dark matter is not constrained by indirect constraints from cosmic rays and Cosmic Microwave Background (CMB) measurements \cite{GMDM,GMDM2}.

When $m_S>m_G$, scalar dark matter can also annihilate into a pair of spin-2 particles through the $t/u$-channels \cite{GMDM,GMDM2,GMDM3,GMDM-dd}, becoming dominant due to sizable spin-2 couplings to dark matter. Then, the corresponding annihilation cross section is given, as follows,
\bea
(\sigma v)_{SS\rightarrow GG} = \frac{4 c_{S}^4 m_{S}^2}{9 \pi \Lambda^4 }
\frac{(1-r_S)^\frac{9}{2}}{r^4_S  (2-r_S)^2}   \label{tch-scalar}
\eea
with $r_S = \left(\frac{m_G}{m_S}\right)^2$.

For light dark matter, the DM annihilations into photons or gluons are relevant. For sub-GeV dark matter, the DM annihilations into mesons must be considered instead of those into gluons.
Then, for scalar dark matter, the annihilation cross sections into a pair of massless gauge bosons \cite{GMDM} are
\bea
(\sigma v)_{SS\rightarrow \gamma\gamma}&\simeq&v^4 \cdot  \frac{c^2_S c^2_\gamma }{60\pi\Lambda^4}\frac{m^6_S}{(4m^2_S-m^2_G)^2+\Gamma^2_G m^2_G},\\
(\sigma v)_{SS\rightarrow gg}&\simeq& v^4 \cdot  \frac{2c^2_S c^2_g }{15\pi\Lambda^4}\frac{m^6_S}{(4m^2_S-m^2_G)^2+\Gamma^2_G m^2_G}.
\eea
For $2m_S\lesssim 1.5\,{\rm GeV}$, instead of the annihilation into a gluon pair, we should consider the annihilation cross section of scalar dark matter into a meson pair, as follows,
\bea
(\sigma v)_{SS\rightarrow \pi\pi} \simeq v^4 \cdot\frac{ c^2_S c^2_\pi}{720\pi \Lambda^4}\, \frac{m^6_S}{(4m^2_S-m^2_G)^2+\Gamma^2_G m^2_G} \left(1-\frac{m^2_\pi}{m^2_S}\right)^\frac{5}{2}
\eea
where $c_\pi\simeq c_q$ in the limit of small momenta of produced pions, because the chiral perturbation theory takes in. 
We also need to include the annihilation of scalar dark matter into charged pions and kaons, if kinematically allowed.

\subsubsection{Fermion dark matter}

The annihilation cross section for fermion dark matter, $\chi{\bar\chi}\rightarrow \psi{\bar\psi}$, is given \cite{GMDM,GMDM2,GMDM3} by 
\bea
(\sigma v)_{\chi{\bar\chi}\rightarrow \psi{\bar\psi}} = v^2 \cdot \frac{N_c c^2_\chi c^2_\psi }{72\pi\Lambda^4}
\frac{m^6_\chi}{(4m^2_\chi-m^2_G)^2+\Gamma^2_G m^2_G} \left(1-\frac{m^2_\psi}{m^2_\chi}\right)^\frac{3}{2} 
\left(3+\frac{2m^2_\psi}{m^2_\chi}\right).
\eea
Thus, the annihilation of fermion dark matter into the SM fermions becomes $p$-wave suppressed. 
Then, similarly to the case of scalar dark matter, fermion dark matter is not constrained by indirect constraints from cosmic rays and CMB measurements \cite{GMDM,GMDM2}. 

When $m_{\chi}>m_G$, fermion dark matter also annihilates into a pair of spin-2 particles through  to the $t/u$-channels \cite{GMDM,GMDM2,GMDM3}, as follows,
\bea
(\sigma v)_{\chi \bar\chi \rightarrow GG} &=& \frac{c_{\chi}^4 m_{\chi}^2}{16 \pi \Lambda^4 }
\frac{(1-r_\chi)^\frac{7}{2}}{r^2_\chi (2-r_\chi)^2}  \label{tch-fermion}
\eea
with $r_\chi = \left(\frac{m_G}{m_\chi}\right)^2$. Then, the resulting annihilation cross section is $s$-wave, so it becomes dominant in determining the relic density for fermion dark matter.

For light fermion dark matter, the annihilation cross sections into a pair of massless gauge bosons and a pair of mesons \cite{GMDM} are
\bea
(\sigma v)_{\chi{\bar\chi}\rightarrow \gamma\gamma}&\simeq& v^2\cdot \frac{c^2_\chi c^2_\gamma}{12\pi\Lambda^4}\frac{m^6_\chi}{(4m^2_\chi-m^2_G)^2+\Gamma^2_G m^2_G},\\
(\sigma v)_{\chi{\bar\chi}\rightarrow gg}&\simeq&  v^2\cdot\frac{2c^2_\chi c^2_g}{3\pi\Lambda^4}\frac{m^6_\chi}{(4m^2_\chi-m^2_G)^2+\Gamma^2_G m^2_G}.
\eea
For $2m_\chi\lesssim 1.5\,{\rm GeV}$, we need to include the annihilation channel into a pion pair by 
\bea
(\sigma v)_{\chi{\bar\chi}\rightarrow \pi\pi}\simeq v^2\cdot \frac{c^2_\chi c^2_\pi }{144\pi\Lambda^4}
\frac{m^6_\chi}{(4m^2_\chi-m^2_G)^2+\Gamma^2_G m^2_G} \left(1-\frac{m^2_\pi}{m^2_\chi}\right)^\frac{5}{2}.
\eea
Similarly, the annihilation of fermion dark matter into charged pions and kaons, if kinematically allowed, should be also included.

\subsubsection{Vector dark matter}

The annihilation cross section for vector dark matter, $XX\rightarrow \psi{\bar\psi}$, is given \cite{GMDM,GMDM2,GMDM3} by
\bea
(\sigma v)_{XX\rightarrow \psi{\bar\psi}}&=& \frac{4N_c c^2_X c^2_\psi }{27\pi \Lambda^4}
 \frac{m^6_X}{(4m^2_X-m^2_G)^2+\Gamma^2_G m^2_G}\left(3+\frac{2m^2_\psi}{m^2_X}\right)\left(1-\frac{m^2_\psi}{m^2_X}\right)^\frac{3}{2}.
\eea
Thus, the annihilation of vector dark matter into quarks becomes $s$-wave. 
In this case, smaller spin-2 mediator couplings to the SM quarks or vector dark matter can be consistent with the correct relic density, as compared to the other cases. 
In this case, the CMB measurement for recombination era can rule out the vector dark matter mass below $100\,{\rm GeV}$, if the relic density is determined solely by the direction annihilation into the SM particles. But, indirect detection signals from the annihilation of vector dark matter are promising \cite{GMDM,GMDM2}.

For $m_X>m_G$,  vector dark matter also annihilates into a pair of spin-2 particles through  the $t/u$-channels \cite{GMDM,GMDM2,GMDM3,GMDM-dd}, as follows,
\bea
(\sigma v)_{X X  \rightarrow GG} &=&
\frac{c_{X}^4 m_{X}^2}{324 \pi \Lambda^4 }
\frac{\sqrt{1-r_X}}{r^4_X  (2-r_X)^2} \,
\bigg(176+192 r_X+1404 r^2_X-3108 r^3_X \nonumber \\
&&+1105 r^4_X+362 r^5_X+34 r^6_X \bigg) 
\label{tch-vector}
\eea
with $r_X = \left(\frac{m_G}{m_X}\right)^2$.

For light vector dark matter, the annihilation cross sections into a pair of massless gauge bosons and a pair of mesons \cite{GMDM} are
\bea
(\sigma v)_{XX\rightarrow \gamma\gamma}&=&\frac{8c^2_X c^2_\gamma}{9\pi\Lambda^4}\frac{m^6_X}{(4m^2_X-m^2_G)^2+\Gamma^2_G m^2_G}, \\
(\sigma v)_{XX\rightarrow gg}&=&\frac{64c^2_X c^2_g}{9\pi\Lambda^4}\frac{m^6_X}{(4m^2_X-m^2_G)^2+\Gamma^2_G m^2_G}.
\eea
For $2m_X\lesssim 1.5\,{\rm GeV}$, we also need to include the annihilation into a pion pair by 
\bea
(\sigma v)_{XX\rightarrow \pi\pi}\simeq \frac{2c^2_X c^2_\pi}{27\pi \Lambda^4}
\frac{m^6_X}{(4m^2_X-m^2_G)^2+\Gamma^2_G m^2_G}\left(1-\frac{m^2_\pi}{m^2_X}\right)^\frac{5}{2}.
\eea
Similarly, the annihilation of vector dark matter into charged pions and kaons, if kinematically allowed, should be also included.

\subsection{Forbidden channels}

When dark matter is lighter than the spin-2 mediator, but their masses are comparable, that is, $m_{\rm DM}\lesssim m_G$, the annihilation of dark matter into a pair of spin-2 particles is forbidden at zero temperature, but it is kinematically allowed due to the tail of the Boltzmann distribution of dark matter at finite temperature, making the so called forbidden channels relevant for determining the DM abundance.
In this subsection, we consider the forbidden channels in association with the spin-2 mediator. 

In the case when the forbidden channels are dominant, the Boltzmann equations, (\ref{Boltzmann-S}) and (\ref{Boltzmann-F}), become
\bea
\dot{n}_\text{DM} + 3Hn_\text{DM}&
\approx &- \langle \sigma v\rangle_{\rm FB}\,  n^2_{\rm DM} \label{Boltzmann3}
\eea
where the forbidden annihilation cross sections are given by
\bea
 \langle \sigma v\rangle_{\rm FB}&\equiv&  \frac{2(n_{G}^\text{eq})^2 }{(n_\text{DM}^\text{eq})^2}\, \langle \sigma v \rangle_{GG \rightarrow {\rm DM\,DM}}  \nonumber \\
 &=& \frac{50}{g^2_{\rm DM}} \, (1+\Delta_G)^3  e^{-2\Delta_G x} \langle\sigma v\rangle_{GG \rightarrow {\rm DM\,DM}}.
\eea
Here,  $\Delta_{G}=(m_{G}-m_{\rm DM})/m_{\rm DM}$, and $g_{\rm DM}$ is the number of degrees of freedom of dark matter, $g_{\rm DM}=1, 4, 3$, for real scalar, Dirac fermion and vector dark matter, respectively. Here, we have used the detailed balance condition for forbidden channels. 
Moreover, for $m_{\rm DM}<m_G$, the cross sections for the inverse annihilation channels are given by 
\bea
 (\sigma v)_{GG\rightarrow SS} &=&{4 c_S^4 m_S^2 \over 225\pi \Lambda^4}{(r_S -1)^{9\over 2}\over r_S^{7/2}}, \\
  (\sigma v)_{GG\rightarrow \chi{\bar\chi}} &=&\frac{4 c_\chi^4 m_\chi^2}{225\pi \Lambda^4}\Big( \frac{r_\chi -1}{r_\chi} \Big)^{7\over 2}(4+3r_\chi), \\
 (\sigma v)_{GG\rightarrow XX} &=&\frac{c_X^4 m_X^2 \sqrt{r_X-1}}{900\pi \Lambda^4 r_X^{7/2}}\,\Big(48-94r_X^2 +106r_X^3+105r_X^4\Big).
\eea

As a result, the relic density for forbidden dark matter \cite{FDM} is given by
\bea
\Omega_{\rm DM} h^2= 5.20\times 10^{-10}\,{\rm GeV}^{-2}   \bigg(\frac{10.75}{g_*} \bigg)^{1/2} \Big(\frac{x_f}{20} \Big)\, e^{2\Delta_{G}x_f} \, h 
\eea
with
\bea
h \equiv\bigg[ \frac{50}{g^2_{\rm DM}}\langle \sigma v\rangle_{GG\rightarrow {\rm DM\,DM}}\, (1+\Delta_G)^3  \Big(1-2\Delta_G x_f\, e^{2\Delta_G x_f}  \int^\infty_{2\Delta_G x_f} dt\, t^{-1}\,e^{-t} \Big)\bigg]^{-1}.
\eea
There is a Boltzmann suppression factor in the effective annihilation cross sections for forbidden channels, so we would need larger couplings of dark matter to the spin-2 mediator for the correct relic density, as compared to the case with allowed $2\rightarrow 2$ channels for $m_{\rm DM}>m_G$.

\subsection{Gravity-mediated $3\rightarrow 2$ processes}

Scalar dark matter can annihilate by $SSS\rightarrow S G$, which can be dominant over the forbidden channels, $SS\rightarrow GG$, for $m_S<m_G<2m_S$. 
Similarly, the $3\rightarrow 2$ processes for fermion dark matter ($\chi{\bar\chi}\chi\rightarrow \chi G$) and vector dark matter ($XXX\rightarrow X G$) can be important for $m_\chi<m_G < 2 m_\chi$ and $m_X< m_G< 2 m_X$, respectively. Thus, we choose $m_{\rm DM}<m_G<2m_{\rm DM}$  in order for the $3\rightarrow 2$ processes to be kinematically open and for the hidden sector $2\rightarrow 2$ annihilations to be forbidden.
In this subsection, we consider the assisted $3\rightarrow 2$ channels with the spin-2 mediator for the first time.

When the $3\rightarrow 2$ annihilation processes are dominant, the Boltzmann equation (\ref{Boltzmann-S}) becomes
\bea
\dot{n}_\text{DM} + 3Hn_\text{DM}
&\approx &-\langle\sigma v^2\rangle_{ 3\rightarrow 2} \, n^3_{\rm DM} \label{Boltzmann2}
\eea
with
\bea
\langle\sigma v^2\rangle_{ 3\rightarrow 2}&\equiv& \left\{ \begin{array}{cc} 2\langle\sigma v^2\rangle_{{\rm DM\,DM\,DM}\rightarrow {\rm DM}\, G},  \quad {\rm DM}=S,\, X, \\  \frac{1}{2}  \langle\sigma v^2\rangle_{\chi{\bar\chi}\chi\rightarrow \chi G}, \quad {\rm DM}=\chi, \end{array}\right.
 \nonumber \\
&\equiv& \frac{\alpha^3_{\rm eff}}{m^5_{\rm DM}}.
\eea
Here, the corresponding $3\rightarrow 2$ annihilation cross sections for scalar and fermion dark matter are
\bea
\langle\sigma v^2\rangle_{SSS\rightarrow SG}&=&{c_S^6 m_S (16-r)^2\sqrt{(16-r)(4-r)} \over 1209323520\pi \Lambda^6 r^4 (r+2)^2 }(7r^3+348r^2-1392r-2176)^2, \\
\langle\sigma v^2\rangle_{\chi\bar{\chi}\chi\rightarrow\chi G}&=&{c_\chi^6 m_\chi (16-r)^3\big((16-r)(4-r)\big)^{3/2} \over 79626240\pi \Lambda^6 r (r+2)^2}.
\eea
As a result, the relic density for SIMP dark matter \cite{FDM,SIMP} is given by
\bea
\Omega_{\rm DM} h^2=1.41\times 10^{-8}\,{\rm GeV}^{-2} \bigg(\frac{10.75}{g_*} \bigg)^{3/4}  \Big(\frac{x_f}{20} \Big)^2 \bigg(\frac{M^{1/3}_P m_{\rm DM} }{\alpha_{\rm eff}} \bigg)^{3/2}. 
\eea

We note that the $3\rightarrow 2$ annihilation cross sections are highly suppressed for perturbative couplings in most of the parameter space, so they are sub-dominant in determining the relic density, as compared to the previously discussed $2\rightarrow 2$ annihilation channels. 
Therefore, we don't consider the SIMP option in the later discussion.

\subsection{Dark matter self-scattering}

Spin-2 mediator can also mediate the self-scattering process of dark matter, in particular, for fermion and vector dark matter, for which there is no renormalizable interaction for self-scattering. 
We can take the gravity-mediated processes to be dominant for dark matter self-scattering and consider the interplay between relic density condition and small-scale problems in galaxies. 

For scalar dark matter, the self-scattering cross section for $SS\rightarrow SS$, divided by DM mass, is in the Born approximation
\bea
\frac{\sigma_{S,{\rm self}}}{m_S}&=& {2 c_S^4 m_S\over 9\pi\Lambda^4 r_S^2} \nonumber \\
&=& 1.5\,{\rm cm^2/g}\,\cdot\bigg(\frac{m_S}{0.1\,{\rm GeV}}\bigg)\bigg(\frac{1\,{\rm GeV}}{\Lambda/c_S}\bigg)^4 \bigg(\frac{m_S}{m_G}\bigg)^4.
\eea

For fermion dark matter, the self-scattering cross section from $\chi{\bar\chi} \rightarrow \chi{\bar\chi}$ and $\chi\chi\rightarrow \chi\chi$ (and its complex conjugate), divided by DM mass are similarly given by
\bea
\frac{\sigma_{\chi,{\rm self}}}{m_\chi}&=& \frac{1}{4m_\chi} (\sigma_{\chi{\bar \chi}}+2\sigma_{\chi\chi}) \nonumber \\
&=& {c_\chi^4 m_\chi \over 18\pi \Lambda^4 r_\chi^2} \nonumber \\
&=&0.39\,{\rm cm^2/g}\,\cdot \bigg(\frac{m_\chi}{0.1\,{\rm GeV}}\bigg)\bigg(\frac{1\,{\rm GeV}}{\Lambda/c_S}\bigg)^4 \bigg(\frac{m_\chi}{m_G}\bigg)^4.
 \eea

Finally, for vector dark matter, the self-scattering cross section for $XX\rightarrow XX$, divided by DM mass, is given by
\bea
\frac{\sigma_{X,{\rm self}}}{m_X}&=& {2 c_X^4 m_X \over 27 \pi \Lambda^4}{(32-56r_X+27r_X^2) \over r_X^2 (4-r_X)^2} \nonumber \\
&=&0.179\,{\rm cm^2/g}\,\cdot \bigg(\frac{m_\chi}{0.1\,{\rm GeV}}\bigg)\bigg(\frac{1\,{\rm GeV}}{\Lambda/c_S}\bigg)^4 \bigg(\frac{m_\chi}{m_G}\bigg)^4\,\, \frac{3(32-56r_X+27r_X^2)}{(4-r_X)^2}.
\eea

We note that for both scalar and fermion dark matter, the DM self-scattering cross section little depends on the DM velocity. In the case of scalar dark matter, there is an $s$-channel contribution with the spin-2 mediator too, but it is velocity-suppressed by the overall factor. 
On the other hand, for vector dark matter, the DM self-scattering cross section could be enhanced at a particular DM velocity due to the $s$-channel resonance \cite{murayama}, so it would be possible to accommodate the velocity-dependent self-interaction, being compatible with galaxy clusters such as Bullet Cluster \cite{bullet}.

\subsection{Unitarity bounds}

As we regard the massive spin-2 particle as a mediator for dark matter in the effective theory, it is important to make a consistency check by unitarity and perturbativity for the spin-2 interactions.
In this subsection, we briefly discuss this issue from dark matter annihilation and self-scattering.

From the DM annihilation cross sections for ${\rm DM\,DM}\rightarrow GG$, given in eqs.~(\ref{tch-scalar}), (\ref{tch-fermion}) and (\ref{tch-vector}), the corresponding scattering amplitudes grow with dark matter in the limit of $r_{\rm DM}=(m_G/m_{\rm DM})^2\lesssim 1$, being bounded by  the partial wave unitarity as follows,
\bea
|{\cal M}_{SS\rightarrow GG}| &\simeq& \frac{8}{3}\, \frac{c^2_S m^2_S}{\Lambda^2}\,\Big(\frac{m_S}{m_G} \Big)^4 <16\pi,  \\
|{\cal M}_{\chi{\bar \chi}\rightarrow GG}| &\simeq& \frac{\sqrt{2}}{2}\, \frac{c^2_\chi m^2_\chi}{\Lambda^2}\,  \Big(\frac{m_\chi}{m_G} \Big)^2 <8\pi, \\
|{\cal M}_{XX\rightarrow GG}| &\simeq& \frac{8 \sqrt{11}}{9}\,  \frac{c^2_\chi m^2_X}{\Lambda^2}\,  \Big(\frac{m_X}{m_G} \Big)^4 <16\pi.
\eea 
Similarly, for dark matter self-scattering, the corresponding scattering amplitudes grow with dark matter mass by $|{\cal M}_{\rm DM\,DM\rightarrow DM\, DM}|\sim \frac{c^2_{\rm DM} m^2_{\rm DM}}{\Lambda^2}\,\Big(\frac{m_{\rm DM}}{m_G} \Big)^2$, so the unitarity bounds from them are less significant for scalar and vector dark matter or comparable for fermion dark matter. Thus, it is sufficient to impose the unitarity bounds from ${\rm DM\,DM}\rightarrow GG$. 

As a result, the unitarity bounds impose the lower bounds on the spin-2 mediator mass depending on the spin of dark matter, as follows,  
\bea
m_G&\gtrsim& 0.48\, \Big(\frac{c_S m_S}{\Lambda} \Big)^{1/2} m_S,  \label{unit1} \\
m_G&\gtrsim& 0.14\,  \Big(\frac{c_\chi m_\chi}{\Lambda} \Big) \, m_\chi,  \label{unit2} \\
m_G&\gtrsim& 0.49\,  \Big(\frac{c_X m_X}{\Lambda} \Big)^{1/2} m_X. \label{unit3}
\eea
Therefore, the case with fermion dark matter is subject to the weakest unitarity bound.
Recently, there is a similar discussion on the unitarity bound on the massive graviton \cite{adam}, based on the Compton scattering process, ${\rm DM}\,G\rightarrow {\rm DM}\, G$, which can set a similar unitarity bound at high energies as for ${\rm DM\,DM}\rightarrow GG$.
In the next section, we take into account the above unitarity bounds in constraining the parameter space with the correct relic density, in particular, for WIMP dark matter.

\section{Detection of dark matter and mediator couplings}

We give the phenomenological discussion on the spin-2 mediator for DM-nucleon elastic scattering, DM-electron elastic scattering, $g-2$ of leptons, meson decays and the direct production at colliders. 
We present for the first time the complete discussion of DM-nucleon scattering in the presence of both quark and gluon couplings and DM-electron scattering as well as the relevance of unitarity at colliders.

\subsection{DM-nucleon elastic scattering}

The scattering amplitude between DM and SM particles through the spin-2 mediator \cite{GMDM-dd} is written in the limit of a small momentum transfer, as follows,
\bea
{\cal M}&=&\frac{ic_{\rm DM} c_{\rm SM}}{2m^2_G\Lambda^2}\, \bigg(2 T^{\rm DM}_{\mu\nu} T^{{\rm SM},\mu\nu} -\frac{2}{3}T^{\rm DM} T^{\rm SM}\bigg)  \nonumber \\
&=&\frac{ic_{\rm DM} c_{\rm SM}}{2m^2_G\Lambda^2}\, \bigg(2{\tilde T}^{\rm DM}_{\mu\nu} {\tilde T}^{{\rm SM},\mu\nu} -\frac{1}{6}{T}^{\rm DM} { T}^{\rm SM}\bigg)
\eea
where ${\tilde T}^{\rm SM(DM)}_{\mu\nu}$ is the traceless part of energy-momentum tensor given by ${\tilde T}^{\rm SM(DM)}_{\mu\nu}=T^{\rm SM(DM)}_{\mu\nu}-\frac{1}{4}\eta_{\mu\nu} {T}^{\rm SM(DM)}$ with ${T}^{\rm SM(DM)}$ being the trace of energy-momentum tensor. 

First, the elastic scattering amplitude between dark matter and nucleon  \cite{GMDM-dd} is given by
\bea
{\cal M}=\frac{ic_{\rm DM} c_{\rm SM}}{2m^2_G\Lambda^2}\,\bigg(2{\tilde T}^{\rm DM}_{\mu\nu} \langle N(p_2) | {\tilde T}^{{\rm SM},\mu\nu} |N(p_1)\rangle -\frac{1}{6}{T}^{\rm DM} \langle N(p_2) | { T}^{\rm SM} |N(p_1)\rangle\bigg).
\eea

For direct detection experiments, we can consider only the contributions from quarks and gluons in a nucleon, as follows, 
\bea
c_{\rm SM}T^{\rm SM}_{\mu\nu}= \sum_q c_q T^q_{\mu\nu} + c_g T^g_{\mu\nu}.
\eea
Then, we get the trace part in the effective theory for three quark flavors ($u,d,s$) and gluons as
\bea
T^{\rm SM} = -\sum_{q=u,d,s}  c_q \bigg[m_q {\bar q} q + \frac{\alpha_S}{12\pi} \, G_{\mu\nu} G^{\mu\nu}\bigg]+\frac{11c_g \alpha_S}{8\pi} \, G_{\mu\nu} G^{\mu\nu},
\eea
where scale anomalies from light quarks and gluons are separately taken into account.
Moreover, the traceless part (twist-2 operators) for five quark flavors ($u,d,s,c,b$) and gluons is given by
\bea
c_{\rm SM} {\tilde T}^{\rm SM}_{\mu\nu} =   \sum_{q=u,d,s,c,b} c_q {\tilde T}^q_{\mu\nu} + c_g {\tilde T}^g_{\mu\nu}.  
\eea
As a result, the nuclear matrix elements for the trace part become
\bea
 \langle N(p) | c_{\rm SM}{ T}^{{\rm SM}} |N(p)\rangle&=&-m_N\bigg[\sum_{q=u,d,s} c_q  \Big(f^{N}_{Tq}-\frac{2}{27} f_{TG}\Big) +\frac{11}{9} c_g  f_{TG}\bigg] {\bar u}_N(p) u_N(p)
 \eea
 where $f^{N}_{Tq}, f_{TG}$ are the mass fractions of light quarks and gluons in a nucleon, respectively, and $ f_{TG}=1-\sum_{q=u,d,s} f^{N}_{Tq}$. 
Here, we used the RG invariant quantity, $ \langle N(p) | \alpha_S G_{\mu\nu} G^{\mu\nu} |N(p)\rangle=-\frac{8\pi}{9}\, f_{TG}m_N$, which is obtained in the effective theory for three quark flavors.
For the universal spin-2 couplings with $c_q=c_g$, we obtain the standard results for 
\bea
 \langle N(p) | { T}^{{\rm SM}} |N(p)\rangle= -m_N \bigg[ \sum_{q=u,d,s} f^{N}_{Tq}+  f_{TG} \bigg] {\bar u}_N(p) u_N(p)=-m_N{\bar u}_N(p) u_N(p).
\eea

On the other hand, the nuclear matrix elements for the traceless part \cite{hisano} are
 \bea
  \langle N(p) | c_{\rm SM}{\tilde T}^{{\rm SM}}_{\mu\nu} |N(p)\rangle&=&\bigg[\sum_{q=u,d,s,c,b} c_q\Big(q(2)+{\bar q}(2) \Big)+c_g G(2)\bigg]  \nonumber \\
  &&\quad\times \frac{1}{m_N}\Big(p_\mu p_\nu - \frac{1}{4} p^2 g_{\mu\nu} \Big) {\bar u}_N(p) u_N(p)
 \eea
 where $q(2), {\bar q}(2)$ and $G(2)$ are the second moments of the parton distribution functions(PDFs) of quark, antiquark and gluon, respectively, 
\bea
q(2)+{\bar q}(2) &=& \int^1_0 dx\, x \,[q(x)+{\bar q}(x)],  \label{pdf2nd} \\
G(2)&=& \int^1_0 dx\, x\, g(x). 
\eea
The mass fractions are  $f^p_{T_u}=0.023$, $f^p_{T_d}=0.032$ and $f^p_{T_s}=0.020$ for a proton and $f^n_{T_u}=0.017$, $f^n_{T_d}=0.041$ and $f^n_{T_s}=0.020$ for a neutron \cite{hisano}. On the other hand, the second moments of PDFs are calulated at the scale $\mu=m_Z$ using the CTEQ parton distribution as $G(2)=0.48$, $u(2)=0.22$, ${\bar u}(2)=0.034$, $d(2)=0.11$, ${\bar d}(2)=0.036$, $s(2)={\bar s}(2)=0.026$, $c(2)={\bar c}(2)=0.019$ and $b(2)={\bar b}(2)=0.012$ \cite{hisano}.

There, using the results in the appendix A, the total cross section for spin-independent elastic scattering between dark matter and nucleus  \cite{GMDM-dd} is given by
\bea
\sigma_{{\rm DM}-A}^{ SI}= \frac{ \mu^2_A}{\pi}\,\Big(Z f^{\rm DM}_p+(A-Z) f^{\rm DM}_n\Big)^2
\eea
where $\mu_A=m_\chi m_A/(m_\chi +m_A)$ is the reduced mass of the DM-nucleus system and
$m_A$ is the target nucleus mass, $Z,A$ are the number of protons and the atomic number, respectively, and the nucleon form factors are given by the same formula for all the spins of dark matter as
\bea
f^{\rm DM}_p&=& \frac{ c_{\rm DM} m_N m_{\rm DM}}{4m^2_G\Lambda^2}\bigg(\sum_{q=u,d,s,c,b}3c_q(q(2)+{\bar q}(2))+3c_g G(2) \nonumber \\
&&+\sum_{q=u,d,s} \frac{1}{3} c_q \Big(f^p_{Tq}-\frac{2}{27} f_{TG} \Big) +\frac{11}{9} c_g f_{TG} \bigg)\nonumber \\
&\equiv&  \frac{ c^p_{\rm eff} c_{\rm DM} m_N m_{\rm DM}}{4m^2_G\Lambda^2}, \label{fp} \\
f^{\rm DM}_n&=& \frac{c_{\rm DM}  m_N m_{\rm DM}}{4m^2_G\Lambda^2}\bigg(\sum_{q=u,d,s,c,b}3c_q(q(2)+{\bar q}(2))+3c_g G(2) \nonumber \\
&&+\sum_{q=u,d,s} \frac{1}{3}c_q \Big(f^n_{Tq}-\frac{2}{27} f_{TG} \Big) +\frac{11}{9} c_g f_{TG} \bigg)\nonumber \\
&\equiv&  \frac{ c^n_{\rm eff} c_{\rm DM} m_N m_{\rm DM}}{4m^2_G\Lambda^2}, \label{fn}  \\
\eea
where ${\rm DM}=\chi, S, X$  for fermion, scalar and vector dark matter, respectively.
Here, as compared to our previous work \cite{GMDM-dd}, we have included the twist-2 gluon operator at tree level as well as loop effects from heavy quarks and gluons in the trace part.

\subsection{DM-electron elastic scattering}

For light dark matter below GeV scale, the DM-nucleon elastic scattering loses the sensitivity for dark matter searches because of the low threshold of the nucleon recoil energy. 
Then, the DM-electron elastic scattering is relevant for direct detection \cite{SIMP}. The corresponding cross sections relevant for direct detection are independent of the spin of dark matter, given by
\bea
\sigma_{{\rm DM}-e}&=&{4 c_e^2 c_{\rm DM}^2 m_e^4m_{\rm DM}^4 \over 9\pi \Lambda^4 m_G^4 (m_e+m_{\rm DM})^2} \nonumber \\
&\approx&1.5\times 10^{-50}\,{\rm cm}^2\,\bigg( \frac{0.5\,{\rm GeV}}{m_{\rm DM}}\bigg)^2 \bigg(\frac{10\,{\rm TeV}}{\Lambda/c_e}\bigg)^2\bigg(\frac{100\,{\rm GeV}}{\Lambda/c_{\rm DM}}\bigg)^2\bigg( \frac{m_{\rm DM}}{m_G}\bigg)^4
\label{DDe}
\eea
where we assumed that $m_{\rm DM}\gg m_e$ in the second line.

Moreover, the graviton mediator should make dark matter remain in kinetic equilibrium \cite{SIMP0,SIMP} during the freeze-out. In this case, independent of the spin of dark matter, the momentum relaxation rate for the kinetic equilibrium of light dark matter is dominated by
\bea
\gamma_{{\rm DM}-e}={127 \pi^5 c_e^2 c_{\rm DM}^2 m_{\rm DM}  \over 270 \Lambda^4 m_G^4 }\, T^8 \label{relax}
\eea
Then, the kinetic equilibrium of dark matter can be achieved during the freeze-out, as far as $\gamma_{{\rm DM}-e}> H\cdot\Big(\frac{m_{\rm DM}}{T}\Big)$ in the case of WIMP dark matter where $H$ is the Hubble expansion parameter, and $\gamma_{{\rm DM}-e}> H\cdot\Big(\frac{m_{\rm DM}}{T}\Big)^2$ in the case of SIMP dark matter \cite{SIMP,DMdyn}.

\subsection{Lepton $g-2$ from the spin-2 mediator}

When the spin-2 mediator couples to leptons, it gives an extra contribution to the anomalous magnetic moment of leptons, as follows \cite{graesser},
\bea
a_l =  \frac{5 c^2_l m^2_l}{16\pi^2\Lambda^2}\, A\Big(\frac{m_l}{m_G} \Big)
\eea
where $A(y)$ is a monotonically decreasing function, given by
\bea
A(y)= \frac{223}{120} - \frac{1}{5}\int^1_0 dx\, \Big(2x y^2 -1\Big)\,\frac{H(x)}{L(x,y)}
-\frac{1}{5} \int^1_0 dx\, \frac{y^2 P(x)}{L(x,y)}
\eea
with $L(x,y)=x^2 y^2 + 1-x $ and
\bea
H(x)&=& x(1-x) \Big(-\frac{28}{3} +\frac{3}{2}x-\frac{1}{2}x^2  \Big), \\
P(x)&=& -\frac{1}{2} x^5 +3x^4 -\frac{44}{3} x^3 + \frac{64}{3} x^2.
\eea
For $m_G\gg m_l$, the loop function $A(x)$ is approximated \cite{hong} to
\bea
A\Big(\frac{m_l}{m_G}\Big)\approx 1+ \bigg(\frac{1}{3}\ln\Big(\frac{m_l}{m_G} \Big) + \frac{11}{72}\bigg) \frac{m^2_l}{m^2_G}\,,
\eea
rendering the $(g-2)_l$ almost independent of the spin-2 mediator mass, as follows,
\bea
a_l\approx 285\times 10^{-11}\,\bigg(\frac{m_l}{m_\mu}\bigg)^2\bigg(\frac{350\,{\rm GeV}}{\Lambda/c_l}\bigg)^2.
\eea

We note that the deviation of the anomalous magnetic moment of muon between experiment and SM
values is given \cite{amu,pdg} by
\be
\Delta a_\mu = a^{\rm exp}_\mu -a^{\rm SM}_\mu = 288(80)\times 10^{-11}, 
\ee
which is a $3.6\sigma$ discrepancy from the SM \cite{pdg}.
Furthermore, there is a $2.4\sigma$ discrepancy reported between the SM prediction for the anomalous magnetic moment of electron and the experimental measurements \cite{ae-exp,ae}, as follows,
\bea
\Delta a_e = a^{\rm exp}_e -a^{\rm SM}_e =-88(36)\times 10^{-14}.
\eea

\subsection{Meson decays}

For a light spin-2 mediator with sub-GeV mass, if coupled to light quarks,  constraints from $K^+\rightarrow \pi^++{\rm invisible}$ \cite{raredecays} can be  relevant. Similarly, $B^+\rightarrow K^++{\rm invisible}$ \cite{bmeson} decays also constrain the spin-2 mediator couplings to quarks similarly.  The current bounds on the branching ratios are given by ${\rm BR}(K^+\rightarrow \pi^++{\rm invisible})<(1.73^{+1.15}_{-1.05})\times10^{-10}$ \cite{raredecays} and ${\rm BR}(B^+\rightarrow K^++{\rm invisible})<1.6\times 10^{-5}$ \cite{bmeson}.
The recent discussion on meson decays in the effective theory for dark matter can be found in Ref.~\cite{tevong}.

The decay width of a down-type quark $q_1$ decaying into another down-type quark $q_2$ and $ G$ is given for $m_G<m_{q_1}$ with $m_{q_2}=0$ \cite{mesondecays}, as follows,
\bea
\Gamma(q_1\rightarrow q_2 G) = \frac{c^2_q G^2_F m^7_{q_1} u(x_1)}{192 (2\pi)^5 \Lambda^2}\bigg|\sum_{f=u,c,t} V_{f1} V^*_{f2} \, v(x_f) \bigg|^2 \label{mesondecay}
\eea
where $V_{f1}$ and $V_{f2}$ are the CKM matrix elements,   $x_1=m^2_G/m^2_{q_1}$, $x_f=m^2_G/m^2_f$, and
\bea
u(x) &=&  (1-x) \Big(1-\frac{3}{2}(x+x^2+x^4) +\frac{7}{2} x^3 \Big),  \\
v(x) &=& \frac{1}{36(x-1)^4} \bigg[44-194 x+243 x^2 -98 x^3+5x^4  \nonumber \\
&&\quad+ 6x(2-15x+10x^2)\log(x) \bigg].
\eea
On the other hand, for $m_G>m_{q_1}$ but $m_{q_1}>2m_{\rm DM}$, we can integrate out the spin-2 mediator, so there exists a three-body decay channel, $q_1\rightarrow q_2+ {\rm DM}+{\rm DM}$, with decay rate  about $\Gamma_3\simeq \frac{c}{16\pi} \frac{m^6_{q_1}}{m^4_G \Lambda^2}\, \Gamma_2$, for $m_{q_1}\gg m_{\rm DM}$, as compared to the two-body decay rate $\Gamma_2$, where $c$ is given from $\Gamma(G\rightarrow {\rm DM\,DM})\simeq c \,m^3_G/\Lambda^2$.

\subsection{Mediator production at colliders}

The massive spin-2 particle can be produced singly from gluon fusion or quark/anti-quark scattering at the LHC, decaying into the SM particles or a pair of dark matter.  
Moreover, in intensity beam or linear colliders, we may also constrain non-universal lepton and photon couplings by the photon energy distribution from $e^+ e^-\rightarrow \gamma\, G$.

First, we obtain the squared amplitude for $e^+ e^-\rightarrow \gamma\, G$, as follows,
\bea
|{\cal M}|^2&=&\frac{e^2 c^2_e}{4 \Lambda^2st(s+t-m_G^2)} \Big(s^2  + 2 t (s + t)- 2 m_G^2 t+m_G^4 \Big) \Big(4 t (s + t) - m_G^2 (s + 4 t) \Big) \nonumber \\
&&+\frac{e^2 c^2_e}{\Lambda^2 s} \Big(\frac{c_\gamma}{c_e}-1\Big)\Big( (s + 2 t)^2 - m_G^2 (s + 4 t)+2 m_G^4 \Big) \nonumber \\
&& 
+\frac{e^2 c^2_e}{6 \Lambda^2 m_G^4 s} \Big(\frac{c_\gamma}{c_e}-1\Big)^2  \bigg\{s^2 (s^2 + 2 s t + 2 t^2)- 2 m_G^2 s (s + t) (s + 6 t)  \nonumber \\
&&\quad + 
   m_G^4 (7 s^2 + 24 s t + 12 t^2) - 12 m_G^6 (s + t) +6 m_G^8 \bigg\}
\eea
where $t=-\frac{1}{2}(s-m_G^2 ) (1-\cos\theta)$.
Therefore, for $c_\gamma=c_e$,  the squared amplitude behaves like $|{\cal M}|^2\sim \frac{s}{\Lambda^2}$ for $s\gg m^2_G$ \cite{unitarity,highpt}, which is expected from the dimension-5 interactions for the spin-2 mediator, $-\frac{1}{\Lambda}\, G_{\mu\nu}T^{\mu\nu}$. However, for $c_\gamma\neq c_e$, the squared amplitude becomes $|{\cal M}|^2\sim \frac{s^3}{m^4_G\Lambda^2}$, which shows that the violation of unitarity at a lower energy. 
A similar phenomenon was observed in the QCD process, $q{\bar q}\rightarrow g\, G$ \cite{unitarity,highpt}, for which $c_g\neq c_q$ would give rise to a similar dependence of the corresponding squared amplitude on the center of mass energy.

For $c_\gamma=c_e$, the production cross section for $e^+e^-\rightarrow \gamma\, G$ with unpolarized electron and positron is given by
\bea
\frac{d\sigma_{\gamma G}}{d\cos\theta} &=& \frac{c_e^2 \alpha}{64\Lambda^2 s^2 (s-m^2_G)^2}\, \bigg[(s-m^2_G)^4(2\cos^4\theta-1)  \nonumber \\
&&\quad +(s^2+m^4_G) \Big(3(s-m^2_G)^2+\frac{4m^2_G s}{\sin^2\theta} \Big) \bigg]. 
\eea
Thus, the angular differential cross section becomes independent of $s$ for $s\gg m_G^2$, as expected from the behavior of the squared amplitude.  A similar conclusion can be drawn also for $q{\bar q}\rightarrow g\, G$  at the LHC.
The above result will be used for imposing the bounds from invisible and visible searches at BaBar in Fig.~\ref{fig:ldm-bound} of the next section.

\subsection{Bounds on WIMP}

\begin{figure}[tbp]
\centering 
\includegraphics[width=.40\textwidth]{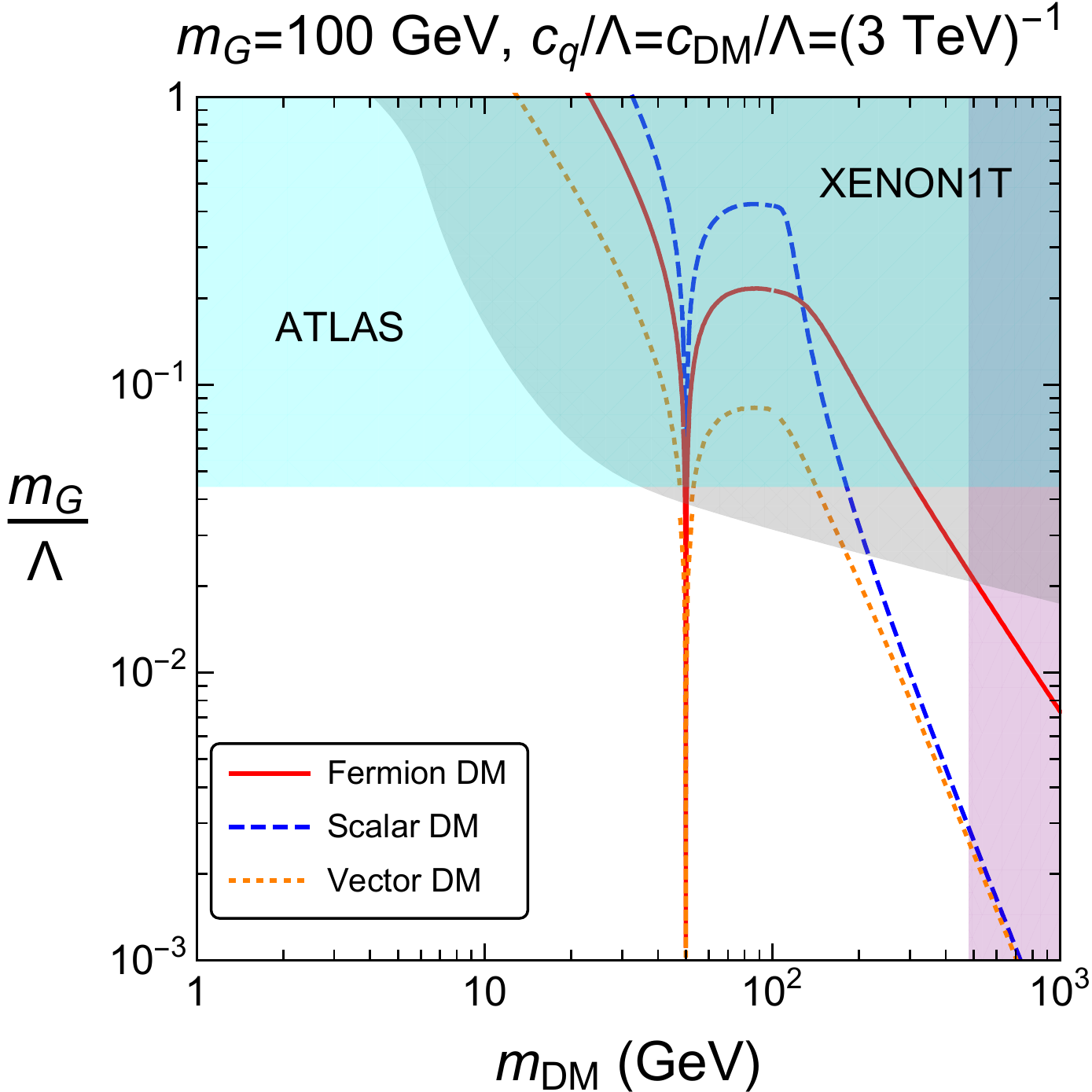} \,\,
\includegraphics[width=.40\textwidth]{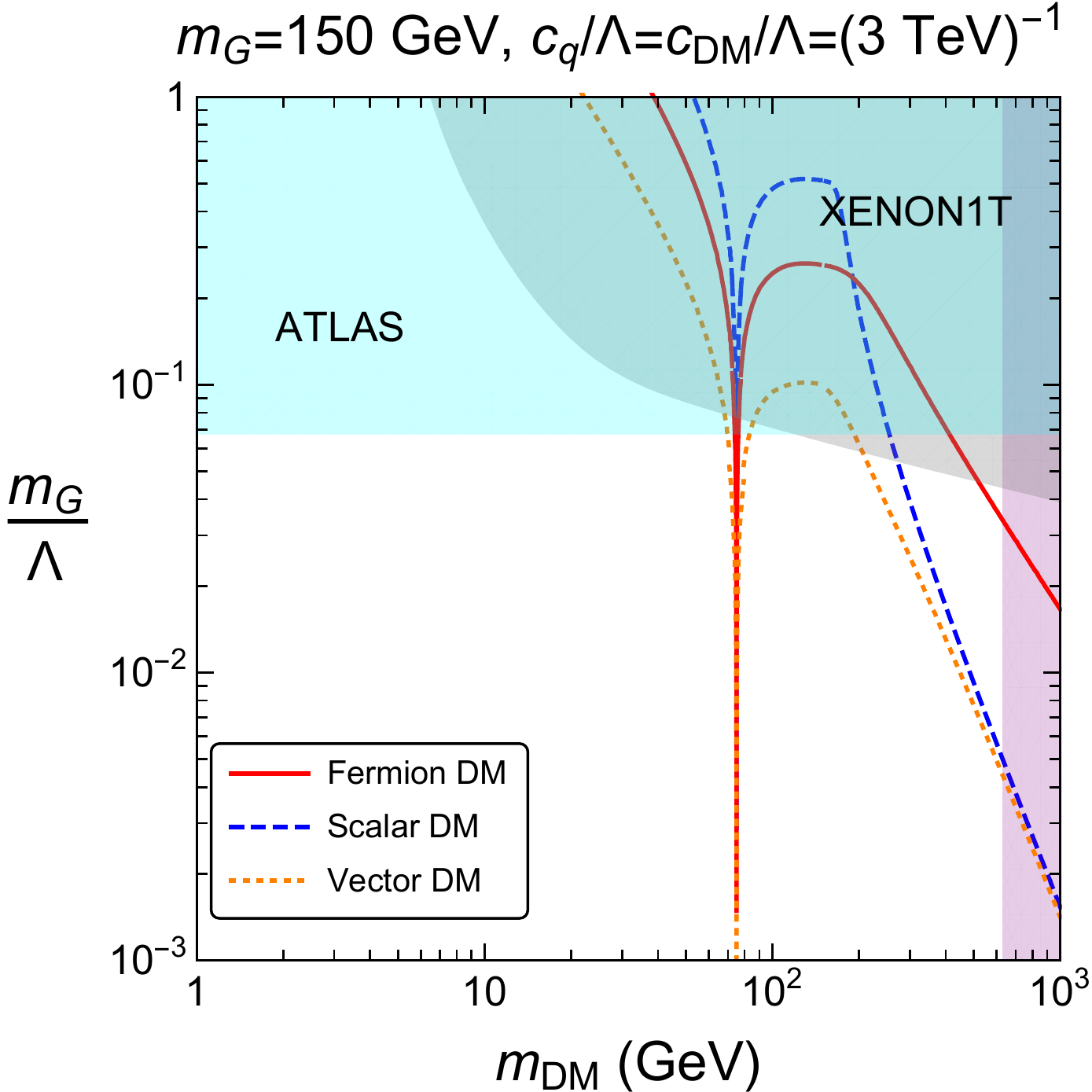} \vspace{0.5cm} \\
\includegraphics[width=.40\textwidth]{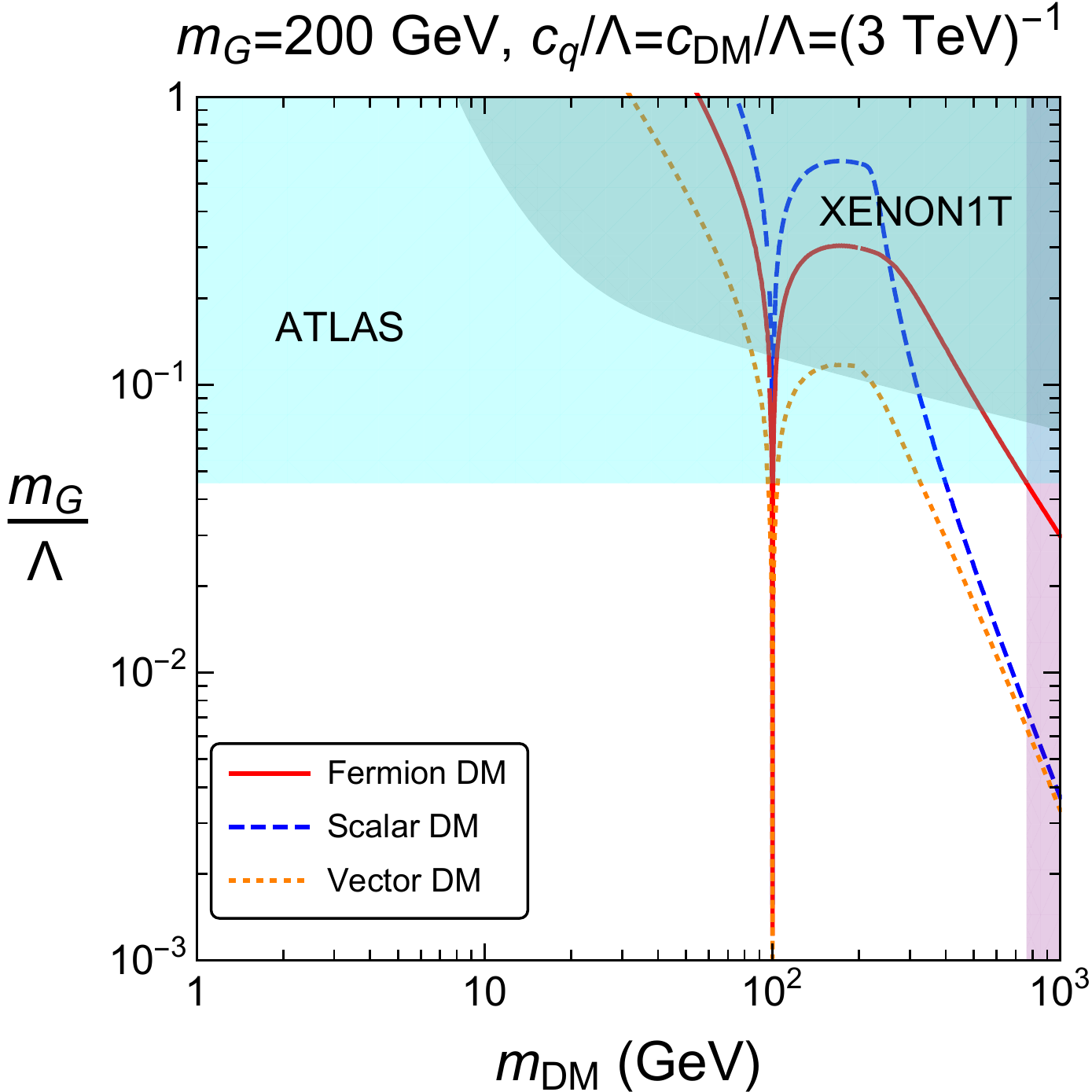}
\caption{\label{fig:wimp1} Parameter space for $m_G/\Lambda$ vs $m_{\rm DM}$ for WIMP dark matter. The relic density is satisfied in red solid, blue dashed and orange dotted lines for fermion, scalar and vector dark matter, respectively. The gray region is excluded by XENON1T and the light blue region is excluded by ATLAS dijet searches.  We have taken the universal spin-2 mediator couplings to the SM and dark matter. The purple region is ruled out by the partial wave unitarity for scalar or vector dark matter. }
\end{figure}

Dijet and dilepton searches at the LHC can constrain relatively heavy spin-2 resonances \cite{heavy-jets}.
Although not sensitive enough, the ISR photon or jet $+$ heavy dijet resonances might be interesting to constrain non-universal quark and gluon couplings by the jet $p_T$ distribution from $q{\bar q}\rightarrow g\, G$ at LHC and future hadron colliders \cite{highpt}.
Direct detection bounds from XENON1T \cite{xenon1t}, LUX \cite{lux}, PandaX \cite{pandax}, etc, are most stringent for weak-scale or heavier dark matter.

For weak-scale spin-2 resonances, the LHC dijet searches are not sensitive due to the large QCD background. Then, dijet resonance $+$ ISR photon \cite{ATLAS-dijet} or jet \cite{ATLAS-jet,CMS-jet} searches can constrain this case. In the presence of dark matter coupling to the spin-2 resonance, the invisible decay of the spin-2 particle with mono-jet of mono-photon is also promising \cite{monojet-invis,DM-forum,DM-forum0,general-dm}.

\begin{figure}[tbp]
\centering 
\includegraphics[width=.45\textwidth]{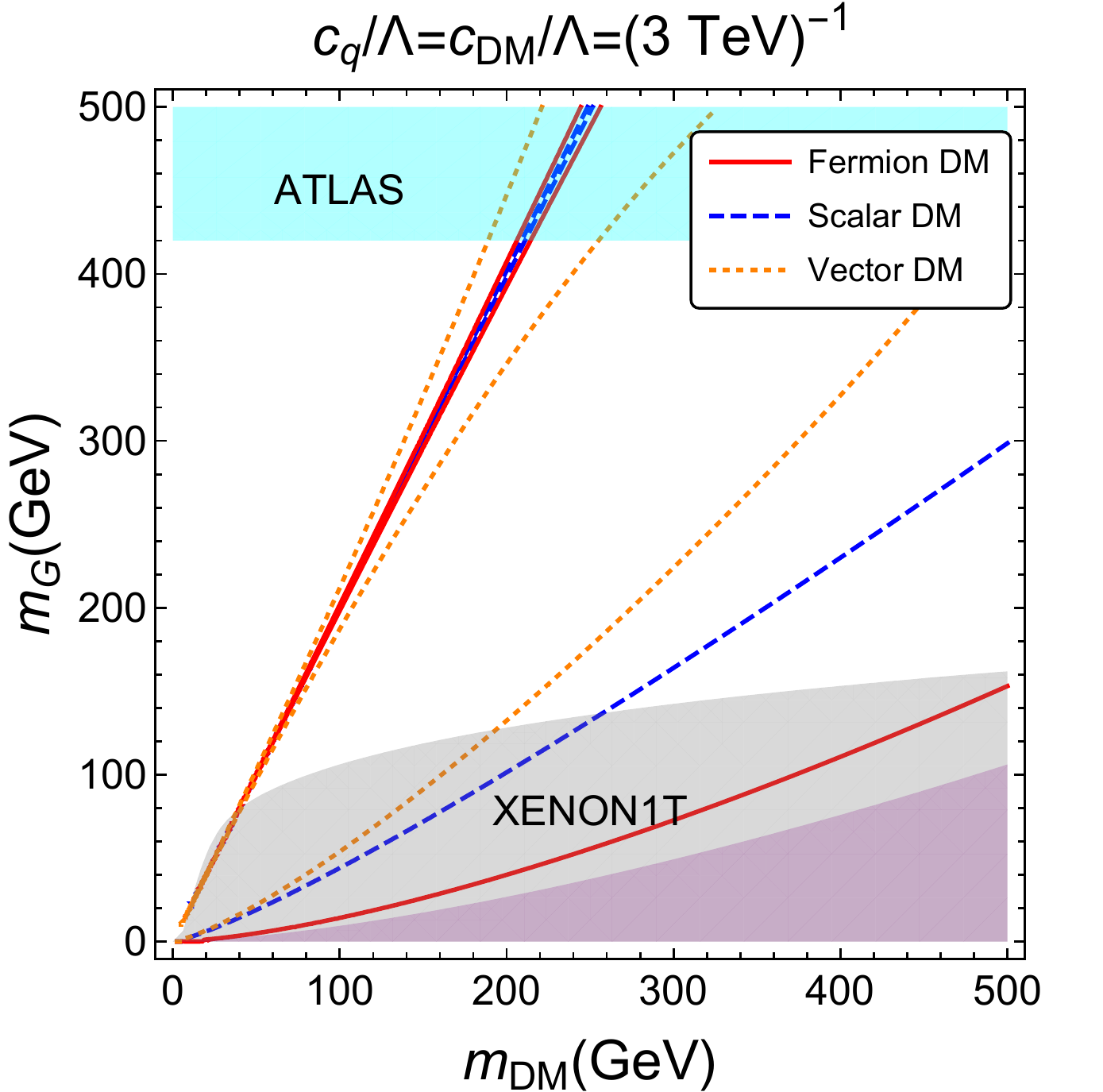} \,\,
\includegraphics[width=.45\textwidth]{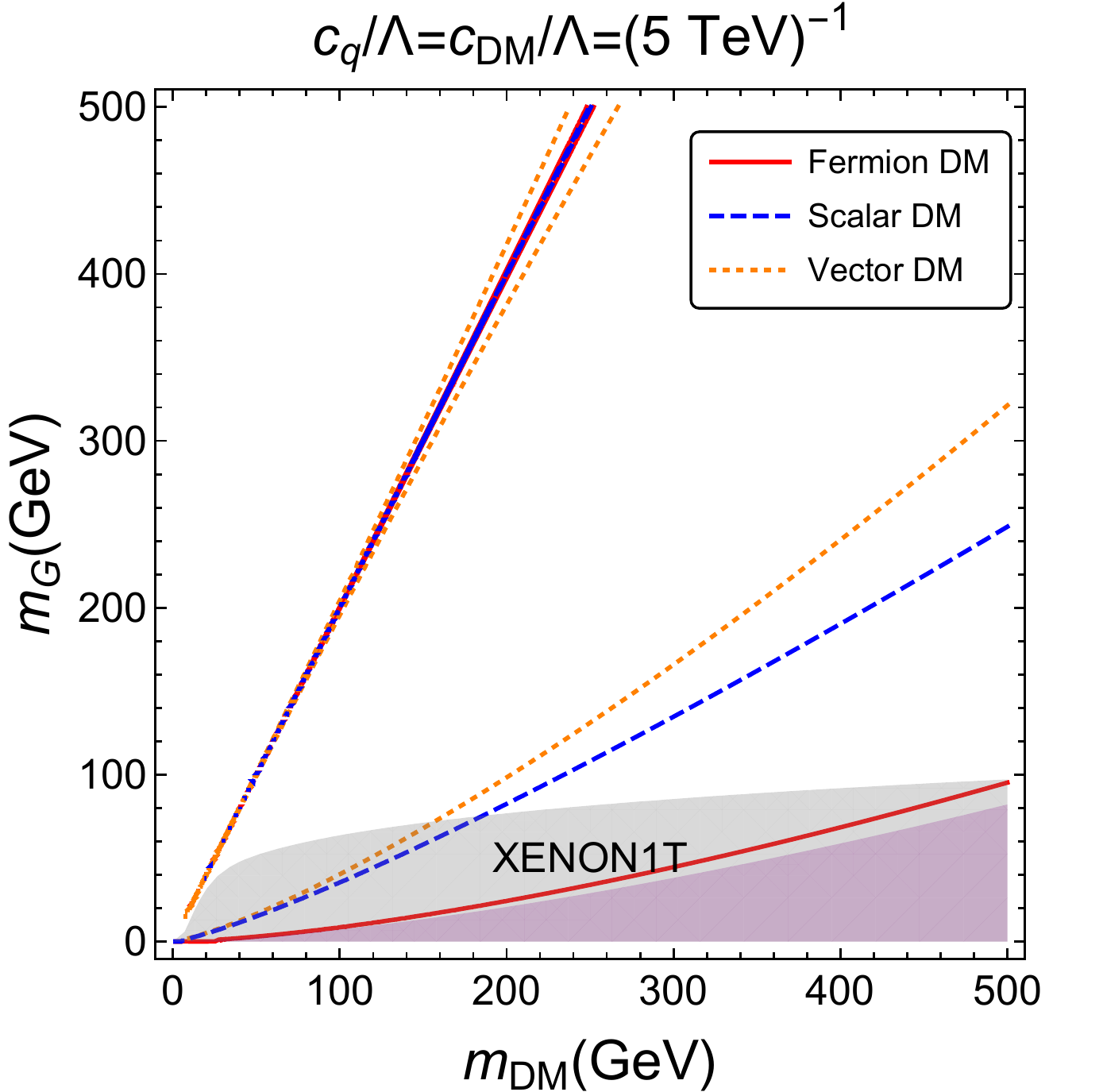} 
\caption{\label{fig:wimp2} Parameter space for  $m_{\rm DM}$ vs $m_G$ for WIMP dark matter. The same as in Fig.~\ref{fig:wimp1}. }
\end{figure}

In Figs.~\ref{fig:wimp1} and \ref{fig:wimp2}, we depict the parameter space for  $m_G/\Lambda$ vs $m_{\rm DM}$ in the former and  $m_{\rm DM}$ vs $m_G$ in the latter, satisfying the correct relic density, in red solid, blue dashed and orange dotted lines for fermion, scalar and vector dark matter, respectively. We took the universal couplings of spin-2 mediator to all the SM quarks and gluons, as well as to dark matter.  
We have excluded the light blue region by the bounds from dijet resonance $+$ ISR photon \cite{ATLAS-dijet} or jet \cite{ATLAS-jet,CMS-jet} searches, and the gray region by the bound on DM-nucleon spin-independent cross section from the direct detection experiment in XENON1T\cite{xenon1t}.  
Moreover, some of the parameter space (in purple) where dark matter is heavier than the spin-2 mediator mass is disfavored by the violation of partial wave unitarity for scalar or vector dark matter as discussed from eqs.~(\ref{unit1})-(\ref{unit3}).  As shown in Fig.~\ref{fig:wimp2}, in a wide parameter space away from the resonance, unitarity constraints turn out to be weaker than the XENON1T bound.

We find from Fig.~\ref{fig:wimp1} that for weak-scale spin-2 mediator, the relic density region below $m_{\rm DM}<m_G/2$ is disfavored by ATLAS dijet bounds.  The XENON1T bound becomes stronger above $m_{\rm DM}>m_G/2$, leaving only the region above $m_{\rm DM}\gtrsim 200\,{\rm GeV}$ or larger masses unconstrained due to the dominance of ${\rm DM\,DM}\rightarrow GG$ channels. But, in this case, the spin-2 mediator produced from the DM annihilation can decay into the SM particles, so the indirect detection experiments from cosmic rays such as positrons, anti-protons and gamma-rays can constrain those large mass regions \cite{GMDM}.
In Fig.~\ref{fig:wimp2},  XENON1T rules out the non-resonance regions below $m_{\rm DM}\simeq 200\,{\rm GeV}$ or $160\,{\rm GeV}$ for the mediator scale, $\Lambda/c_q=3, \, 5\,{\rm TeV}$, but leaves the resonance regions with $m_G=2m_{\rm DM}$ untouched.

\subsection{Bounds on light dark matter}

In the case of light dark matter, we would need a light spin-2 mediator in order to make the  annihilation cross section of dark matter sufficiently large. In this case, monophoton $+$ leptons at BaBar \cite{LM-vis}, and missing energy at BaBar \cite{LM-inv-babar}, Belle-2 \cite{LM-inv-belle2p,LM-inv-belle2}, LHCb (for $m_G>10\,{\rm GeV}$) \cite{lhcb} as well as beam dump experiments such as E137 in SLAC \cite{beamdump}, N64 in CERN SPS \cite{NA64}, etc, can be important to constrain the light spin-2 mediator couplings, in particular, the couplings to leptons and dark matter.
There are also direct detection bounds on DM-electron scattering from XENON10 \cite{xenon10}, DarkSide-50 \cite{darkside}, Sensei experiments \cite{sensei}, etc.

\begin{figure}[tbp]
\centering 
\includegraphics[width=.45\textwidth]{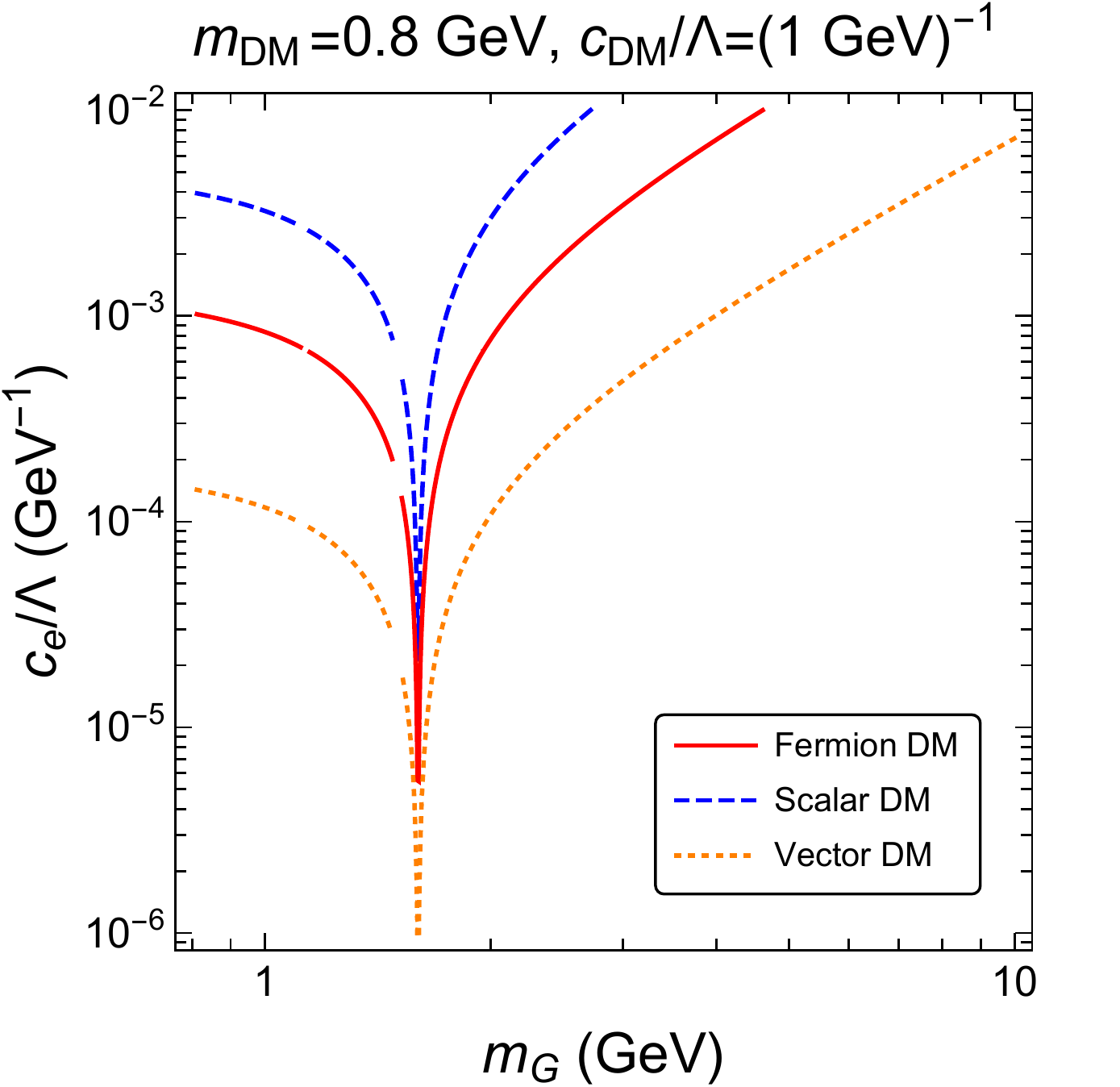}\,\,
\includegraphics[width=.45\textwidth]{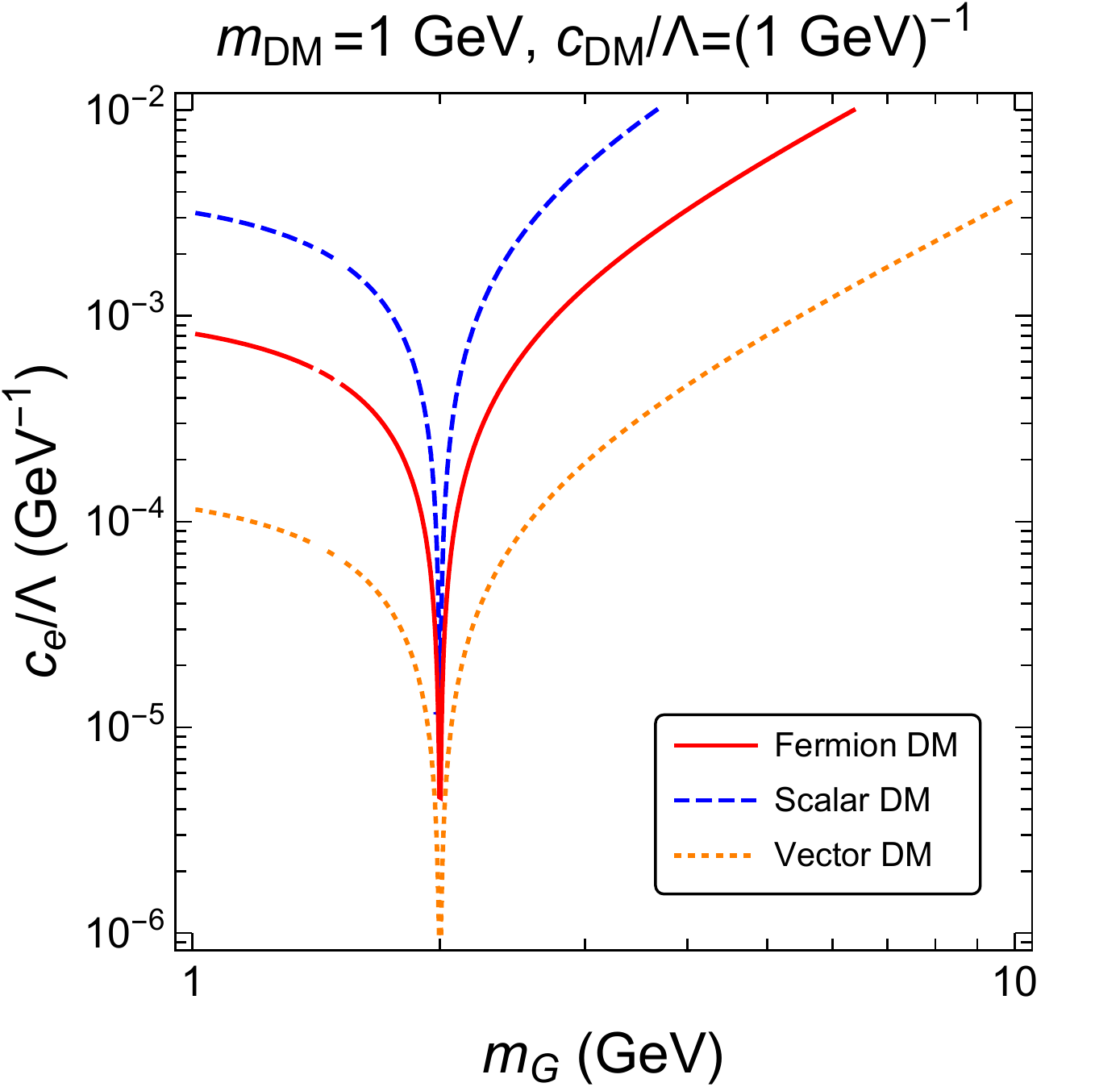}
\caption{\label{fig:sdm1} Parameter space for $c_e/\Lambda$ vs $m_G$ for light dark matter with $m_{\rm DM}<m_G$.  The correct relic density is satisfied in red solid, blue dashed and orange dotted lines for fermion, scalar and vector dark matter, respectively. We chose $m_{\rm DM}=0.8, 1\,{\rm GeV}$ on left and right plots, respectively, and $c_{\rm DM}/\Lambda=(1\,{\rm GeV})^{-1}$ for both plots. }
\end{figure}

For a light spin-2 mediator, we can consider the bounds from $\gamma\,+$ missing energy \cite{LM-inv-babar} or leptons \cite{LM-vis} at BaBar experiment. For the former case, the cosine of the scattering angle of the photon in the center of mass frame was chosen to $|\cos\theta^*_\gamma|<0.6$, and the center of mass energy was $\sqrt{s}=10.58\,{\rm GeV}$. 
Then, we get the limit on the lepton couplings for $m_G<8\,{\rm GeV}$ from invisible and visible searches at BaBar, respectively, as follows,
\bea
\frac{c_e}{\Lambda} &<& 2\times 10^{-4}\,{\rm GeV}^{-1}, \qquad {\rm BaBar\,\, invisible}, 
\\
\frac{c_e}{\Lambda} &<& 3\times 10^{-5}\,{\rm GeV}^{-1}, \qquad {\rm BaBar\,\, visible}. 
\eea
Here, we assumed ${\rm BR}(G\rightarrow {\rm DM\, DM})=1$ in the former and ${\rm BR}(G\rightarrow l{\bar l})=1$ in the latter. So, in general, the above bounds scale up by $1/\sqrt{\rm BR}$. 
The above limits, in particular, from the invisible searches, will be improved by a factor of three in the lepton couplings in Belle-2 experiment \cite{LM-inv-belle2p,LM-inv-belle2}. 

We remark that if we took non-universal couplings by $c_\gamma\neq c_e$, the above bounds from BaBar would become stronger, due to the growth of the corresponding cross section.

Moreover, if the spin-2 mediator is much lighter than $K$-meson or $B$-meson, we can approximate the above partial decay rate of a flavor-changing down-type quark from eq.(\ref{mesondecay}) to
\bea
\Gamma(q_1\rightarrow q_2 G)\approx  \frac{121}{497664\pi^5}\, \frac{c^2_q G^2_Fm^7_{q_1} }{\Lambda^2}\, (V_{t1} V^*_{t2})^2. 
\eea
Therefore, from the current limits on the invisible decays of $K^+$ or $B^+$, we can put the bound on the quark couplings as
\bea
\frac{c_q}{\Lambda} &<& 0.3\,{\rm GeV}^{-1}, \quad\quad\qquad\quad K^+\rightarrow \pi^++{\rm invisible}, \label{meson1} \\
\frac{c_q}{\Lambda} & <& 1.8\times 10^{-2}\,{\rm GeV}^{-1}, \quad \quad B^+\rightarrow K^++{\rm invisible}. \label{meson2}
\eea
As a result, the bounds on quark couplings from meson decays are relatively weaker than those on lepton couplings from BaBar as will be shown in the above.
When the spin-2 mediator is heavier than mesons but dark matter is light enough, mesons can still decay invisibly into a pair of dark matter \cite{tevong}. But, in this case, the bounds on quark couplings become much weaker because of the phase-space suppression for three-body decays of mesons.

\begin{figure}[tbp]
\centering 
\includegraphics[width=.45\textwidth]{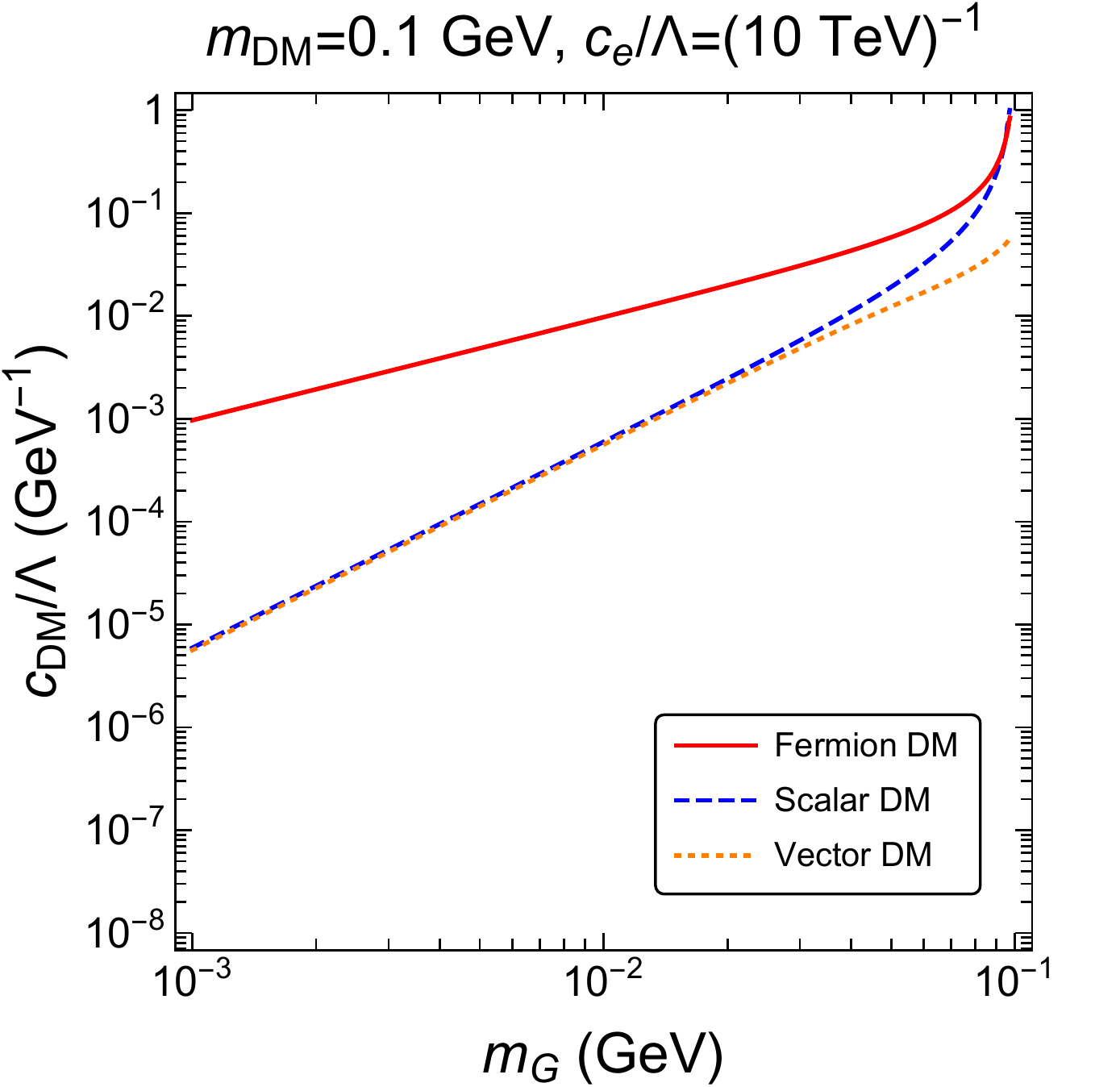}\,\,
\includegraphics[width=.45\textwidth]{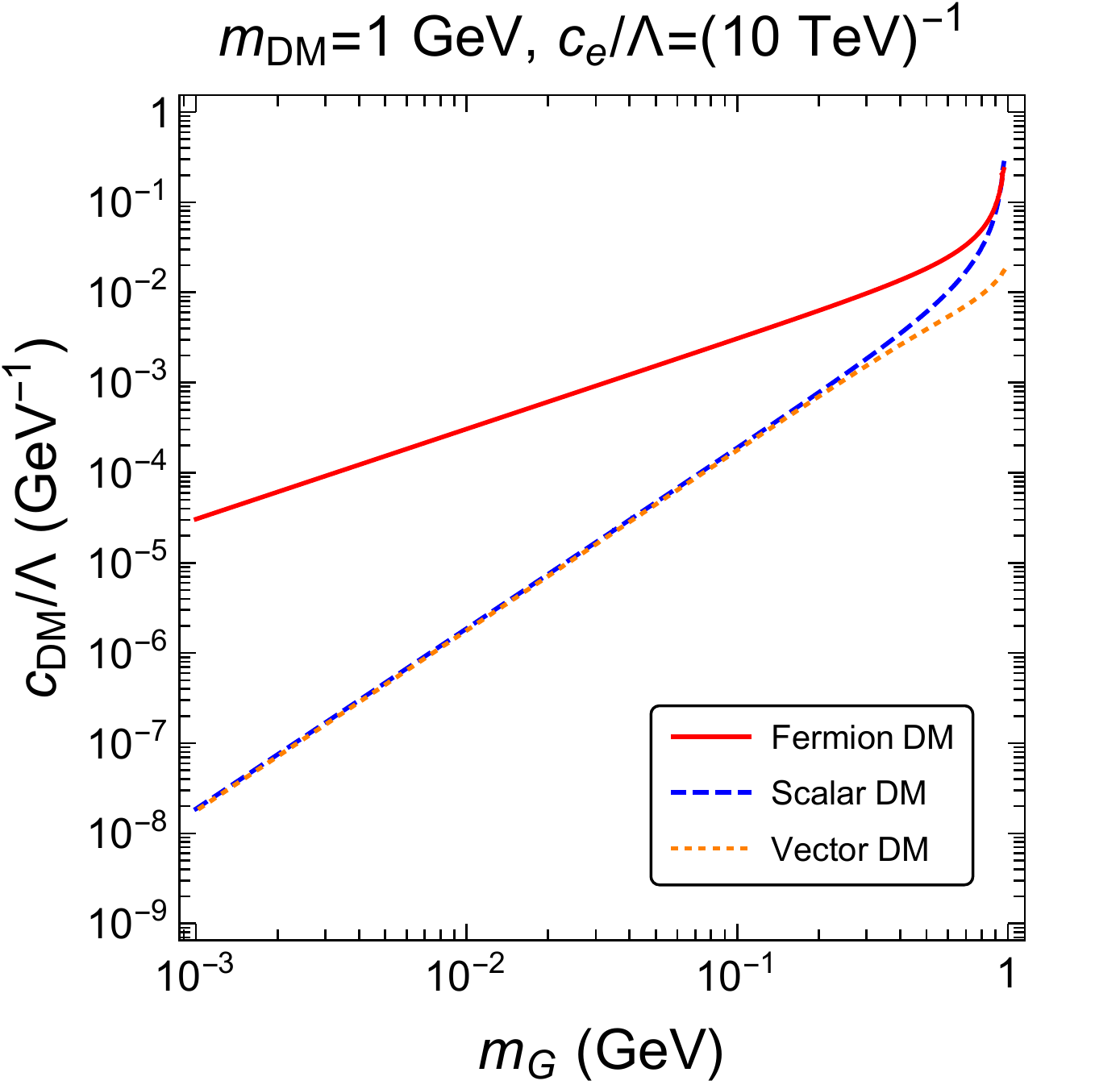}
\caption{\label{fig:sdm2} Parameter space for $c_{\rm DM}/\Lambda$ vs $m_G$ for light dark matter  with $m_{\rm DM}>m_G$. The correct relic density is satisfied in red solid, blue dashed and orange dotted lines for fermion, scalar and vector dark matter, respectively.  We took $m_{\rm DM}=0.1, 1\,{\rm GeV}$ on left and right plots, respectively, and $c_e/\Lambda=(10\,{\rm TeV})^{-1}$ for both plots. }
\end{figure}

\begin{figure}[tbp]
\centering 
\includegraphics[width=.45\textwidth]{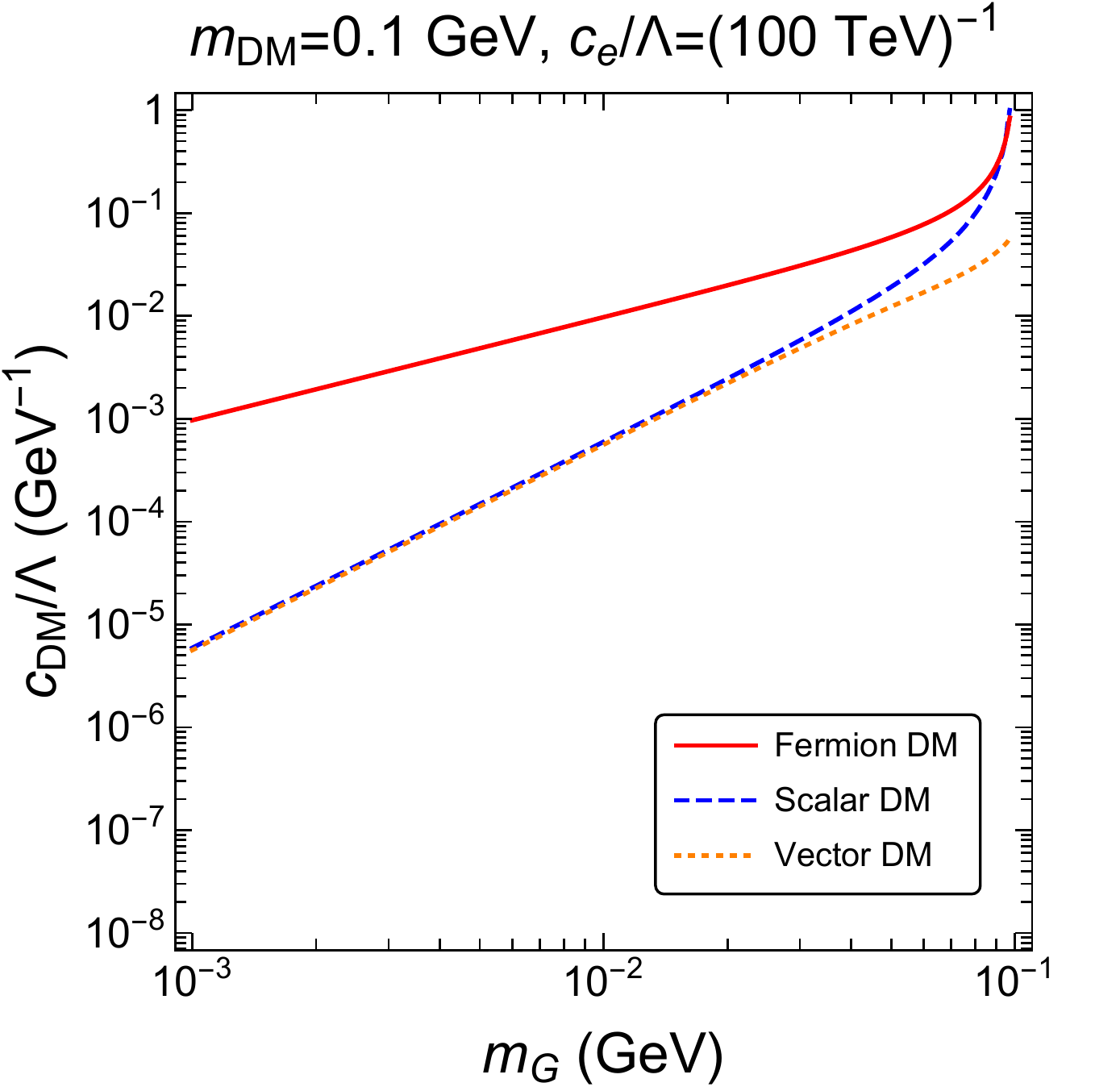}\,\,
\includegraphics[width=.45\textwidth]{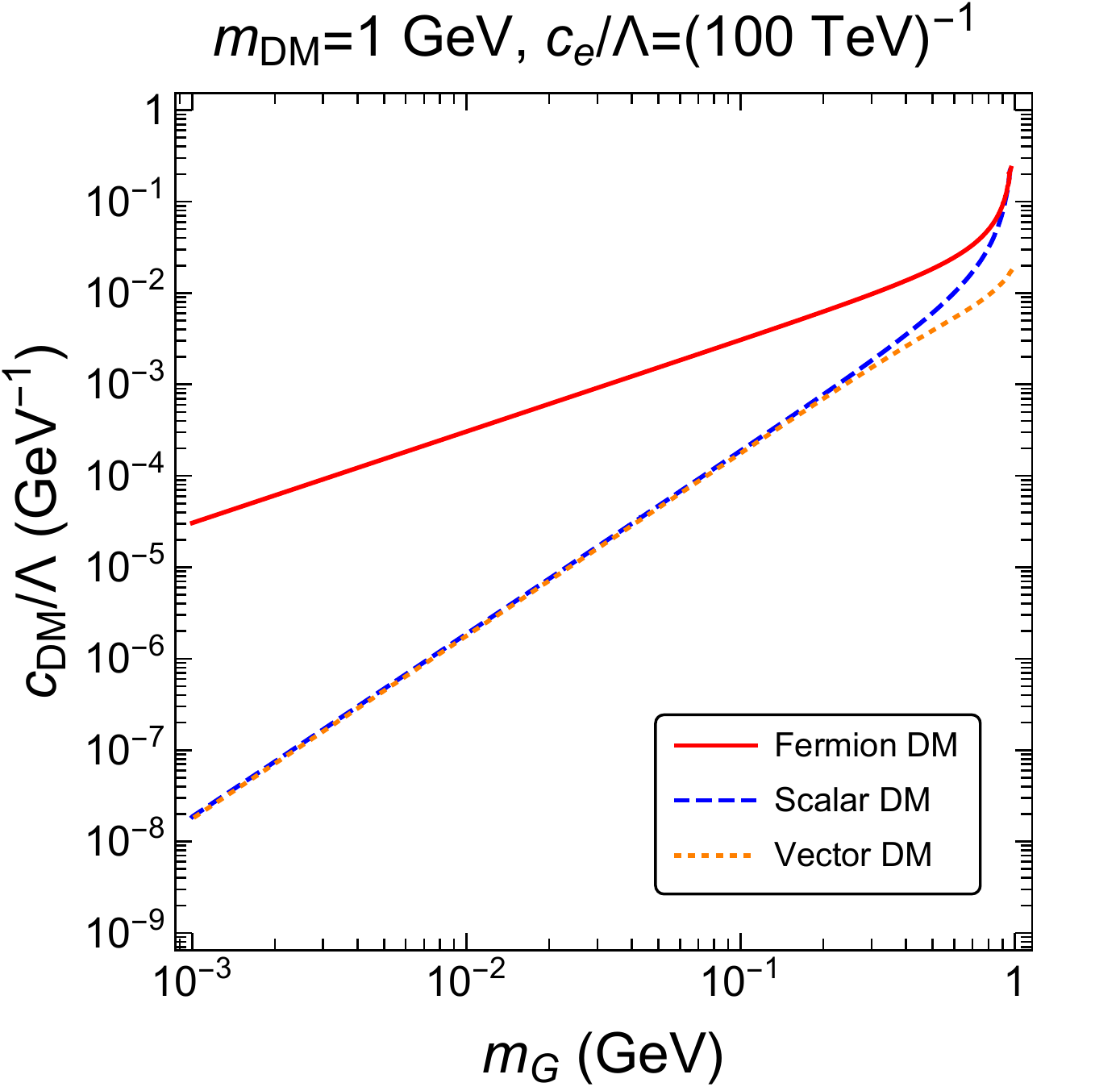}
\caption{\label{fig:sdm3} The same as in Fig.~\ref{fig:sdm2}, except for $c_e/\Lambda=(100\,{\rm TeV})^{-1}$. }
\end{figure}

In Figs.~\ref{fig:sdm1}, \ref{fig:sdm2} and \ref{fig:sdm3}, we show the parameter space for light dark matter below the GeV scale mass satisfying the correct relic density, in $c_e/\Lambda$ vs $m_G$ in the former and  $c_{\rm DM}/\Lambda$ vs $m_G$ in the latter two. For Fig.~\ref{fig:sdm1}, we took $m_{\rm DM}<m_G$ such that dark matter annihilates only into the SM particles, not into a pair of spin-2 mediators. In this case, we find that the graviton couplings to the SM particles satisfying the correct relic density would be strongly constrained by BaBar and other intensity experiments, except the region near the resonance.   On the other hand, for Figs.~\ref{fig:sdm2} and \ref{fig:sdm3}, we took $m_{\rm DM}>m_G$ for which dark matter can annihilate into a pair of spin-2 mediators. In this case, even for a small graviton coupling to the SM particles, for instance, for $\Lambda/c_e=10\,{\rm TeV}$ or $100\,{\rm TeV}$ in  Figs.~\ref{fig:sdm2} or \ref{fig:sdm3}, for which the current experimental constraints are satisfied, we can achieve the correct relic density in a wide range of parameter space for dark matter coupling and spin-2 mediator mass. We note that the DM annihilation into a pair of spin-2 mediators is $s$-wave, so the spin-2 mediators produced from the DM annihilation decay into the SM particles and inject energy into electrons and photons, affecting the CMB recombination \cite{planck}. But, the spin-2 mediator can couple very weakly to the SM, being still consistent with a correct relic density, such that it is long-lived at least as long as the era of the CMB recombination.

We have also checked in Figs.~\ref{fig:sdm1}, \ref{fig:sdm2} and \ref{fig:sdm3} that the DM self-scattering cross sections in the parameter space explaining the relic density are much below $\sigma_{\rm self}/m_{\rm DM}=1\,{\rm cm}^2/g$, the Bullet cluster bound \cite{bullet}.  We also noted that the unitarity bounds given in eqs.~(\ref{unit1})-(\ref{unit3}) are satisfied in the parameter space of the plots in Figs.~\ref{fig:sdm1}, \ref{fig:sdm2} and \ref{fig:sdm3}.

We note that the difference between DM ($c_{\rm DM}/\Lambda$) and lepton couplings ($c_e/\Lambda$) can be explained by the localization of dark matter and leptons in different positions of the extra dimension.
For instance, in RS model, light dark matter can be localized on the IR brane with a small IR scale whereas the SM leptons are localized towards the UV brane \cite{GMDM}.

\begin{figure}[tbp]
\centering 
\includegraphics[width=.45\textwidth]{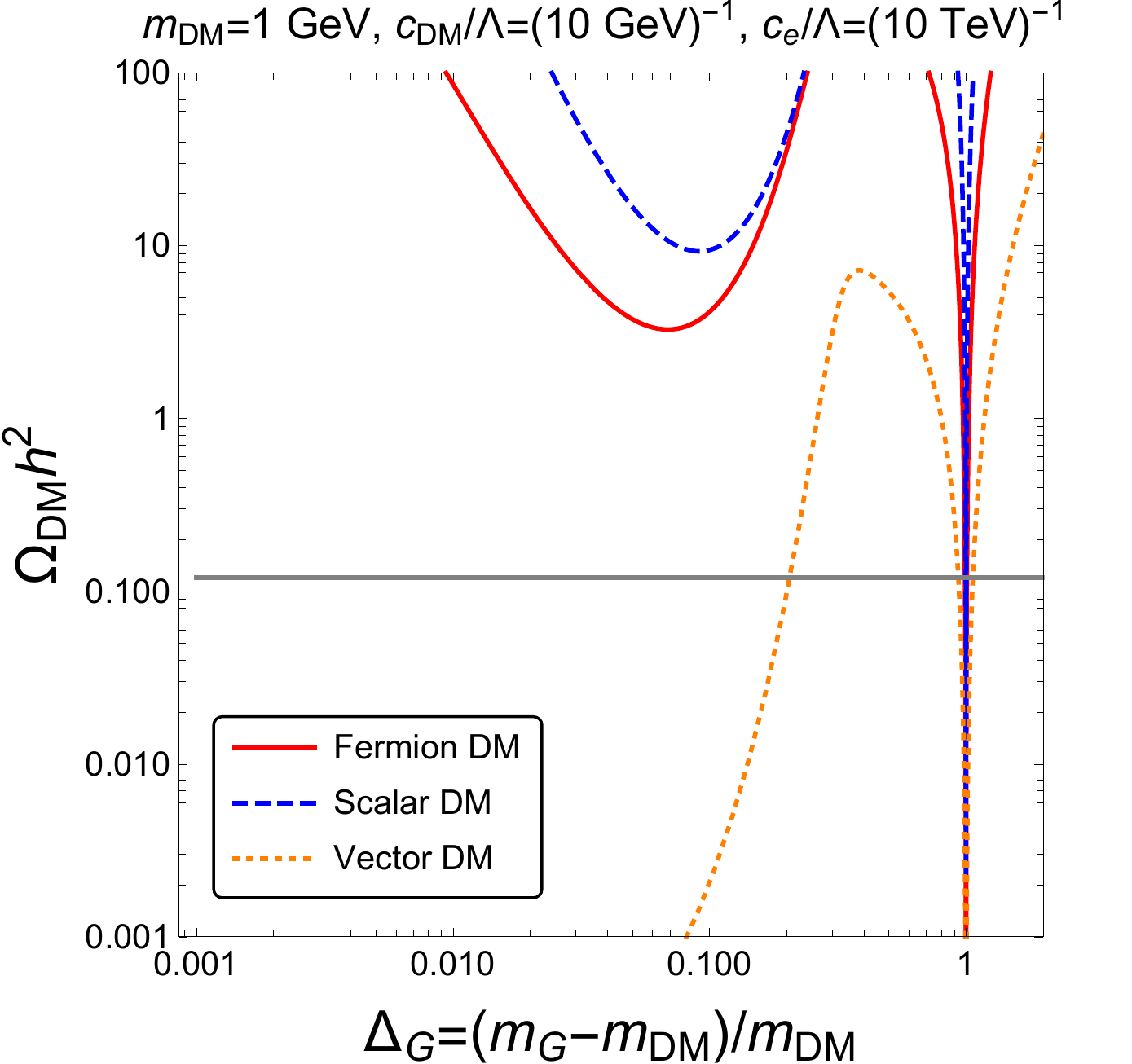}\,\,
\includegraphics[width=.45\textwidth]{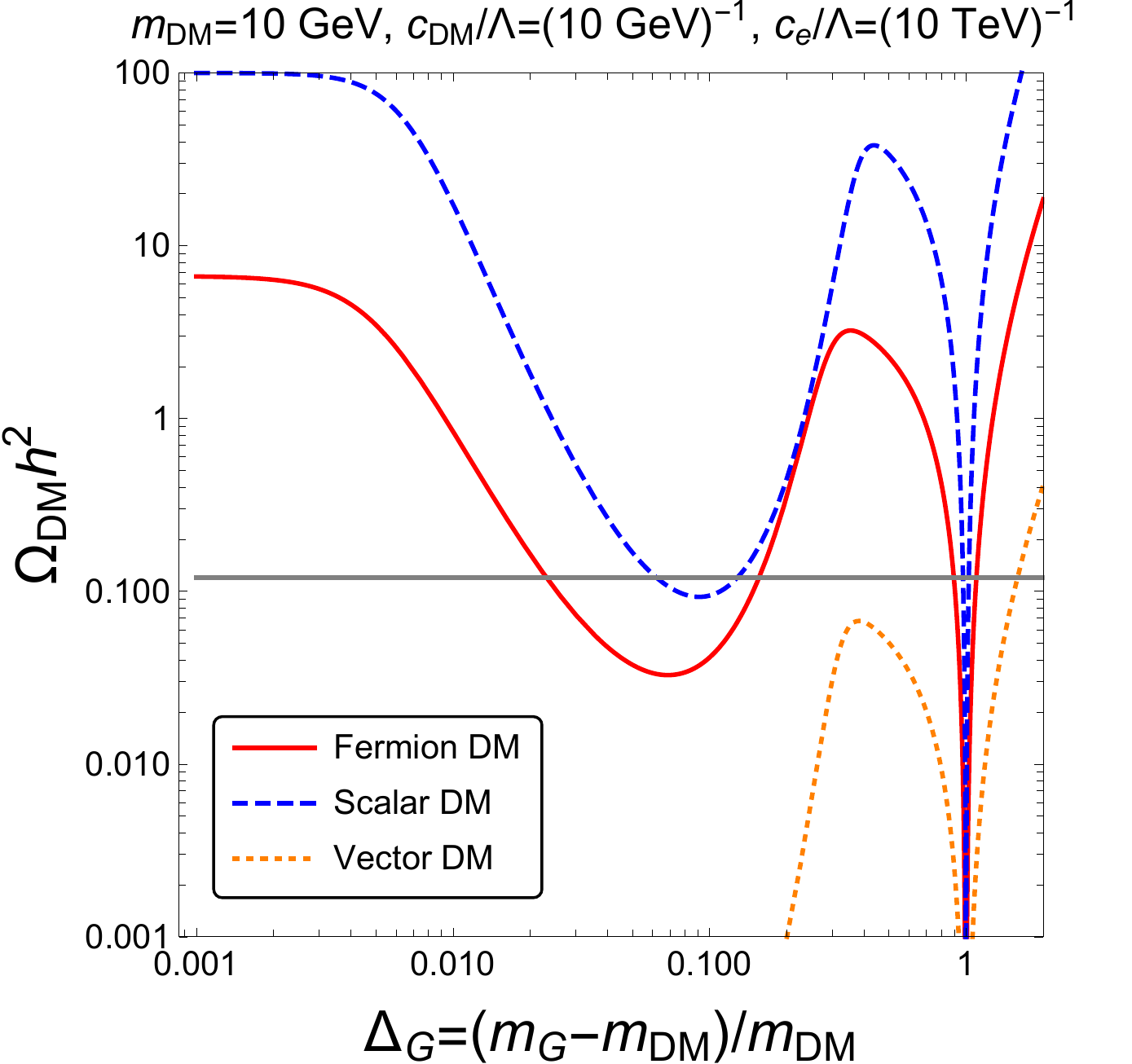}
\caption{\label{fig:sdm4} Relic density as a function of $\Delta_G\equiv (m_G-m_{\rm DM})/m_{\rm DM}$ with forbidden channels included.  The correct relic density is satisfied in red solid, blue dashed and orange dotted lines for fermion, scalar and vector dark matter, respectively.  We took $m_{\rm DM}=1, 10\,{\rm GeV}$ on left and right plots, respectively, and $c_{\rm DM}=(10\,{\rm GeV})^{-1}$ and $c_e/\Lambda=(10\,{\rm TeV})^{-1}$ for both plots.  }
\end{figure}

\begin{figure}[tbp]
\centering 
\includegraphics[width=.45\textwidth]{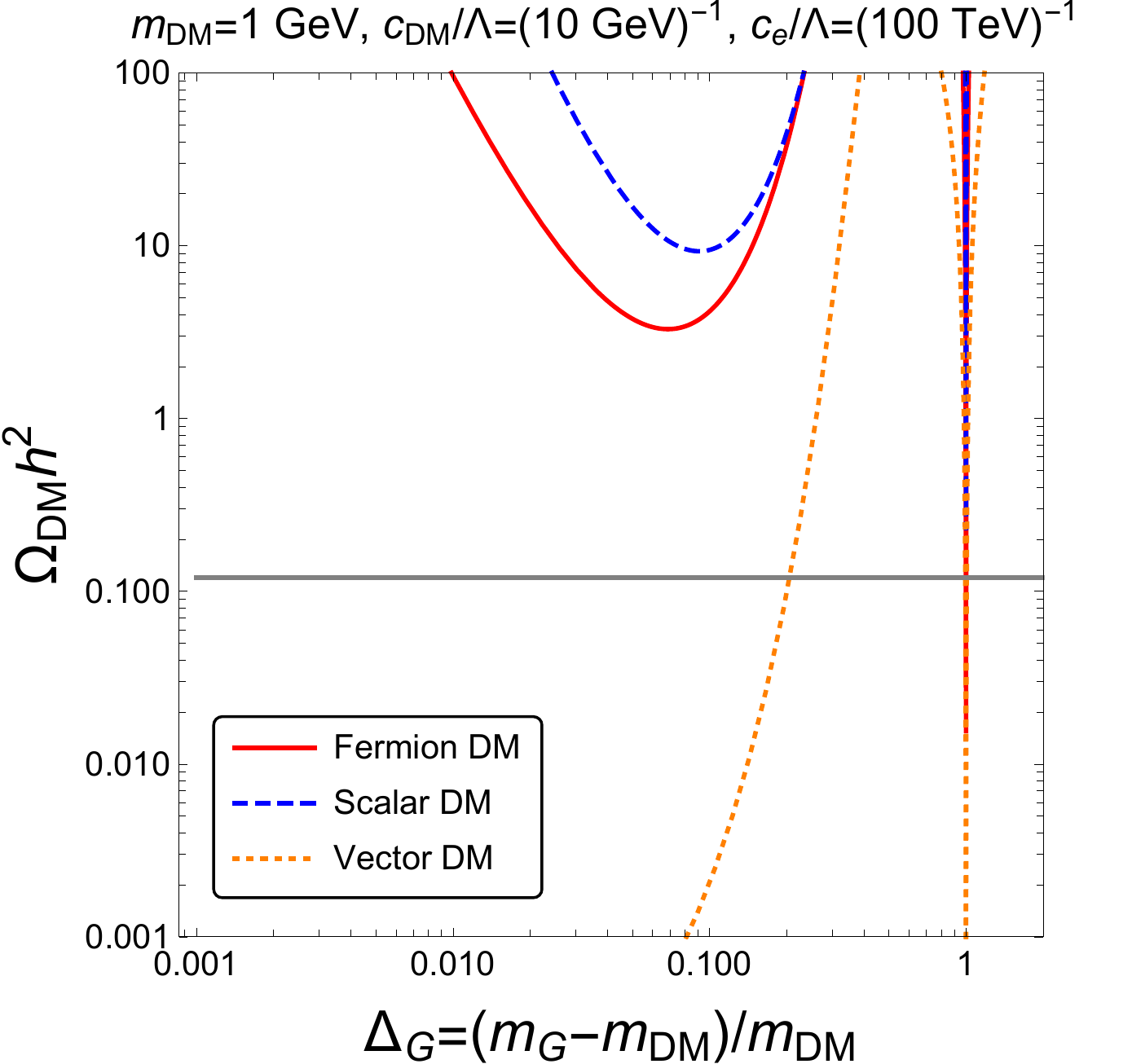}\,\,
\includegraphics[width=.45\textwidth]{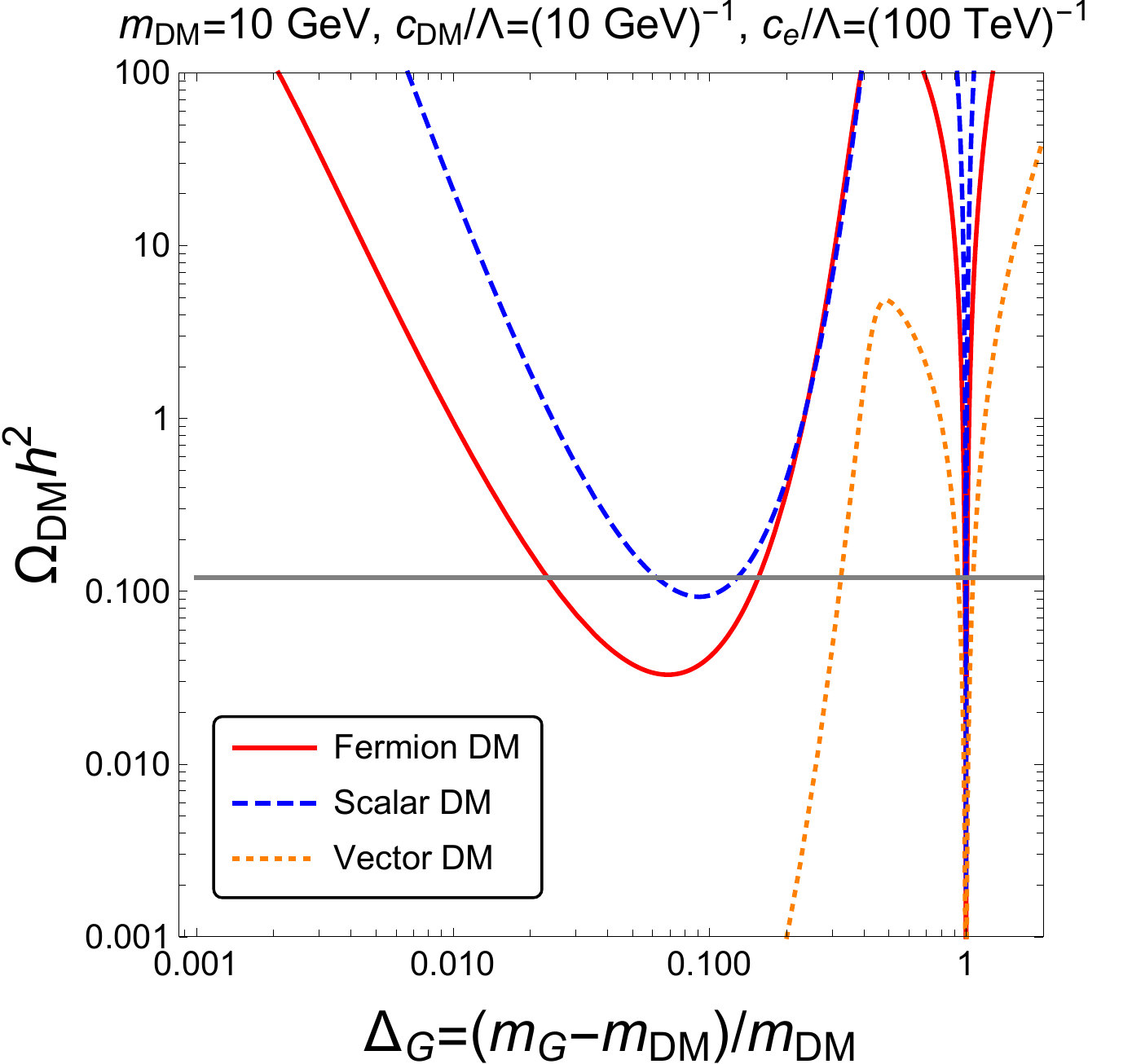}
\caption{\label{fig:sdm5}  The same as in Fig.~\ref{fig:sdm4}, except for $c_e/\Lambda=(100\,{\rm TeV})^{-1}$. }
\end{figure}

In Figs.~\ref{fig:sdm4} and \ref{fig:sdm5}, we present the relic density as a function of the mass difference, $\Delta_G\equiv (m_G-m_{\rm DM})/m_{\rm DM}$, with forbidden channels included. These plots illustrate the role of the forbidden channels in determining the relic density for the spin-2 mediator slightly heavier than dark matter.  In this case, the annihilation of dark matter into a pair of spin-2 mediators is possible only at a nonzero temperature, thus leading to a Boltzmann suppression factor for the corresponding annihilation cross section. For each  of Figs.~\ref{fig:sdm4} and \ref{fig:sdm5}, we have chosen $m_{\rm DM}=1,\,10\,{\rm GeV}$ on left and right. We took $\Lambda/c_{\rm DM}=10\,{\rm GeV}$ for both, and $\Lambda/c_e=10\,{\rm TeV}, \, 100\,{\rm TeV}$ for Figs.~\ref{fig:sdm4} and \ref{fig:sdm5}, respectively.

We find that the correct relic density for vector dark matter can be obtained with smaller couplings to the spin-2 mediator and sub-GeV DM massses, due to a mild phase-space suppression for $m_G\gtrsim m_{\rm DM}$.  On the other hand, for scalar or fermion dark matter, dark matter masses should be about $10\,{\rm GeV}$ or larger for the correct relic density being consistent with perturbativity,  due to significant phase-space suppressions for $m_G\gtrsim m_{\rm DM}$.  The forbidden channels are $s$-wave but get suppressed as the velocity of dark matter decreases in the later stage of the universe and in local galaxies. Thus, the forbidden channels are safe from the indirect bounds from cosmic rays or CMB recombination. In particular, it is remarkable that sub-GeV vector dark matter with $m_{\rm DM}\lesssim m_G$ can be consistent with both the relic density and indirect detection bounds, being compatible with perturbativity.

\begin{figure}[tbp]
\centering 
\includegraphics[width=.45\textwidth]{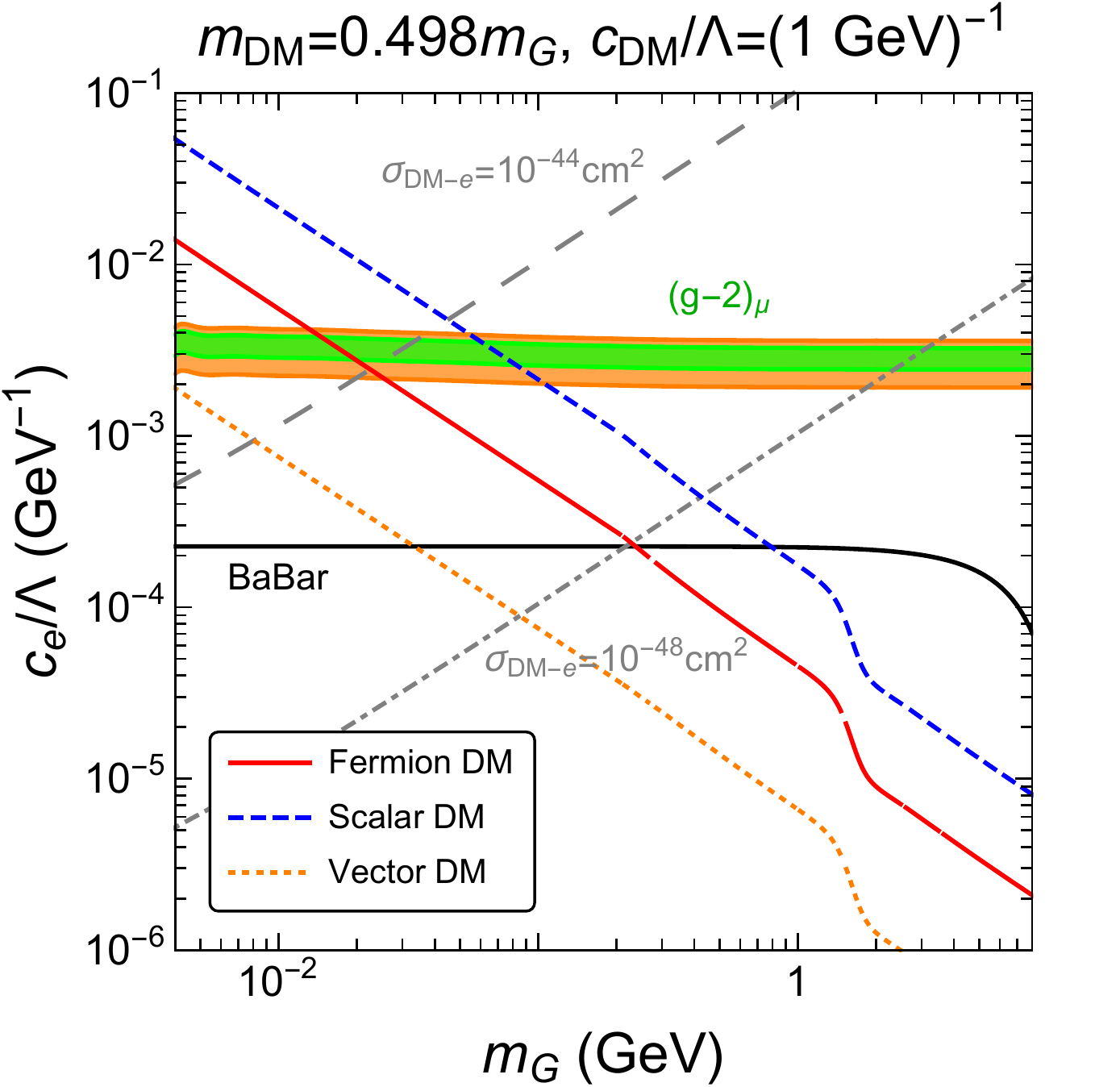}\,\,
\includegraphics[width=.45\textwidth]{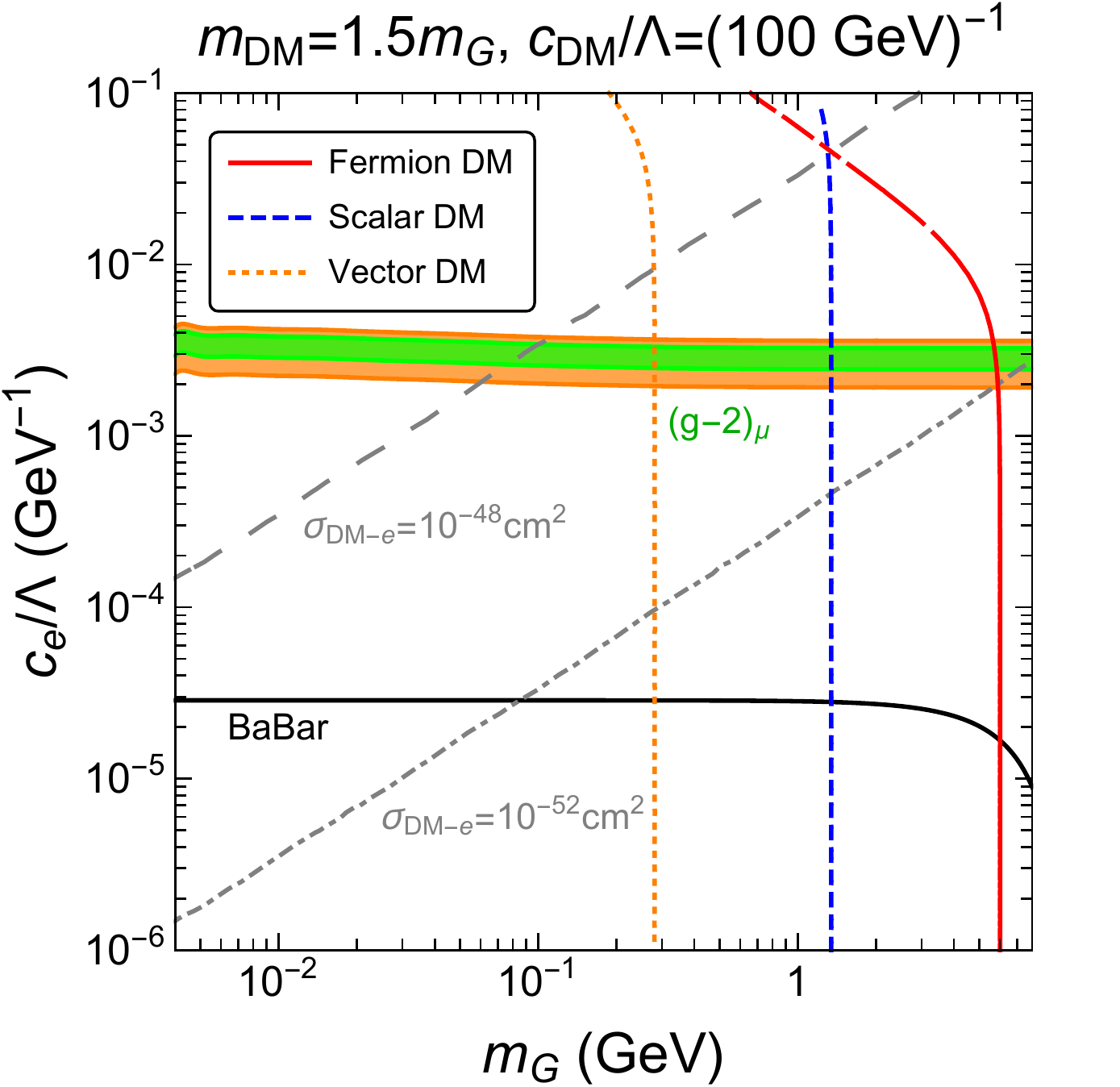}
\caption{\label{fig:ldm-bound} Experimental constraints on $c_e/\Lambda$ vs $m_G$. The correct relic density is obtained along  in red solid, blue dashed and orange dotted lines for fermion, scalar and vector dark matter, respectively. The $(g-2)_\mu$ favored region at $1\sigma$ or $2\sigma$ is shown in green and orange, respectively. Invisible and visible searches at BaBar rule out the region above the black lines on left and right, respectively. Contours for DM-electron scattering cross sections are shown in gray lines for $\sigma_{\rm DM-e}=10^{-44}, \,10^{-48}\,{\rm cm}^2$ on left and $\sigma_{\rm DM-e}=10^{-48},\, 10^{-52}\,{\rm cm}^2$ on right.  }
\end{figure}

In Fig.~\ref{fig:ldm-bound}, we impose various experimental constraints and theoretical constraints in the parameter space for $c_e/\Lambda$ vs $m_G$. We chose the spin-2 mediator mass and dark matter coupling as $m_G=m_{\rm DM}/0.498$ and $\Lambda/c_{\rm DM}=1\,{\rm GeV}$ on left and $m_G=m_{\rm DM}/1.5$ and $\Lambda/c_{\rm DM}=100\,{\rm GeV}$ on right. 
We note that for both plots of  Fig.~\ref{fig:ldm-bound}, the DM self-scattering cross sections in the parameter space of our interest are well below the Bullet cluster bound.

In the left plot of Fig.~\ref{fig:ldm-bound}, the spin-2 mediator can decay dominantly into a pair of dark matter in most of the parameter space satisfying the relic density shown in red solid, blue dashed and orange dotted lines for fermion, scalar and vector dark matter, respectively.  So, the bound from invisible searches at BaBar applies to the whole parameter space below $m_G=8\,{\rm GeV}$, excluding the relic density region for scalar dark matter below $m_G=0.8\,{\rm GeV}$ but less constraining the counterparts for fermion or vector dark matter. The future Belle-2 results \cite{LM-inv-belle2p,LM-inv-belle2}  could improve the limits or probe the larger portion of the relic density regions. 
We also show the $(g-2)_\mu$ favored region in green and orange at $1\sigma$ and $2\sigma$ levels, respectively, but it is excluded by BaBar for the universal lepton couplings \footnote{For $c_e\ll c_\mu$, however, we can make the $(g-2)_\mu$ favored region compatible with the bounds from BaBar. This is possible if leptons are localized at different locations in the warped extra dimension. }. In the same plot, we show the gray contours for DM-electron scattering cross section with $\sigma_{\rm DM-e}=10^{-44}, \,10^{-48}\,{\rm cm}^2$, but most of the parameter space survives the current direct detection bounds on light dark matter, such as XENON10, DarkSide-50, Sensei experiments. We note that as shown in the results, (\ref{meson1}) and (\ref{meson2}), the bounds from $K^+\rightarrow \pi^++G$ or $B^+\rightarrow K^++G$ with $G\rightarrow {\rm invisible}$ are much weaker than BaBar invisible searches, so they are not shown in Fig.~\ref{fig:ldm-bound}.

On the other hand, in the right plot of Fig.~\ref{fig:ldm-bound}, as shown in Figs.~\ref{fig:sdm2} and \ref{fig:sdm3}, we don't need large graviton couplings to the SM particles in the region with $m_{\rm DM}>m_G$, because dark matter can annihilate directly into a pair of spin-2 mediators.   
Therefore, the relic density can be determined almost independent of the graviton couplings to the SM particles, so a lot of parameter space for the correct relic density can be compatible with the current experiments. In this case, the spin-2 mediator decays only into the SM particles, so mono-photon $+$ leptons at BaBar applies, limiting the lepton couplings to the spin-2 mediator. In the same plot, we also show the gray contours for DM-electron scattering cross section with $\sigma_{\rm DM-e}=10^{-48}, \,10^{-52}\,{\rm cm}^2$, so most of the parameter space is unconstrained by direct detection yet. 
We also noted that the unitarity bounds given in eqs.~(\ref{unit1})-(\ref{unit3}) are satisfied in the parameter space of the plots in Fig.~\ref{fig:ldm-bound}.

\section{Spin-2 mediators from the warped extra dimension}

We can regard the spin-2 mediator as the first Kaluza-Klein(KK) mode of graviton from the warped extra dimension or a composite state in a dual conformal field theory.
In the case of the warped extra dimension, there are heavier Kaluza-Klein(KK) modes of graviton, which can be summed up to modify the DM processes, such as DM annihilation and scattering.

After compactification of the warped extra dimension, in principle, nonzero cubic self-couplings for KK gravitons appear in the low energy and they could contribute to the calculations of DM annihilation and scattering processes. 
As the initial 5D gravity theory with the warped extra dimension is ghost-free, there must be no ghost problem in the resulting 4D effective gravity theory. 
The quadratic and cubic self-couplings for KK gravitons are also present in the 4D effective theory and they could change the calculations of the DM annihilations, ${\rm DM\,DM}\rightarrow GG$. 
Moreover, the complete analysis at the non-linear level with cubic self-couplings for KK gravitons would be also relevant for constructing a consistent model of the massive spin-2 particle without a ghost problem at the non-linear level \cite{dRGT,dRGT-review,bimetric}, and showing the delayed violation of unitarity to a higher energy \cite{adam}, and pinning down the UV nature of spin-2 mediators.
Related to the above issue, there are attempts to make a consistent framework without ghosts for a massive spin-2 particle with self-interactions in the literature in the context of massive gravity \cite{dRGT-review} or bi-gravity \cite{bimetric}.

In this work, we didn't attempt to tackle the detailed calculations of the DM annihilations, ${\rm DM\,DM}\rightarrow GG$ with KK gravitons, or the ghost problem of a massive spin-2 particle at the non-linear level. Instead, we assumed that there are only five physical degrees of freedom for a massive spin-2 particle and introduced the interactions of the massive spin-2 particle to matter in the form of energy-momentum tensors. We took the Pauli-Fierz mass term for a massive spin-2 particle and its matter couplings at the linear level, so there is no issue of ghost problem at this level. The mass term for a massive spin-2 particle leads to the non-conservation of energy-momentum tensor, being proportional to the mass term, which is attributed to the breakdown of translational invariance in the warped extra dimension or conformal symmetry in a dual field theory.

In this section, 
motivated by two benchmark models with the warped extra dimension that will be described later, we keep only the linear couplings for a tower of KK gravitons and discuss the impacts of those KK gravitons on DM $s$-channel annihilations into the SM particles and DM scattering processes. For this, only the linear couplings for KK gravitons are sufficient for our discussion. 
We first summarize the KK graviton masses and couplings for two benchmark models with the warped extra dimension and discuss the effects of the heavier KK modes in determining the relic density, the direct detection bounds as well as the direct production of KK gravitons at colliders, in order.
In the end of the section, we remark on the impacts of nonlinear interactions of spin-2 mediators and the unitarity constraint on the DM annihilations, ${\rm DM\,DM}\rightarrow GG$, and discuss those issues  in the ghost-free realization of the massive spin-2 particle.

\subsection{Spin-2 mediator masses and couplings}

The KK modes of graviton in Randall-Sundrum(RS) model \cite{rs} are spaced almost equally. So, if dark matter is lighter than almost twice the mass of the first KK mode, the heavier KK modes would not change much our discussion with the first KK mode only. Otherwise, we need to include the heavy KK resonances explicitly. 
On the other hand, in the 5D continuum limit of the clockwork model, so called the linear dilaton model \cite{dilaton,clockwork0,CW-hmlee, clockwork}, the KK modes of graviton are almost degenerate with a mass gap from the zero mode, challenging for experimental tests \cite{clockwork-pheno,CMS-clockwork}.
So, it is crucial to include the heavier KK modes in the DM processes in this case. 

Suppose that $m_n$ are KK graviton masses, and $c_{{\rm DM},n}, c_{{\rm SM},n}$ are the couplings of the $n$th KK mode to dark matter and the SM, respectively, and depending on the localization in the extra dimension. Here, dark matter and the SM particles can be localized on the IR brane, in which case dark matter has sizable couplings to the SM particles.  But, when the SM particles are localized away from the IR brane, we can just rescale $c_{{\rm SM},n}$ to small values. 

 In the case where dark matter and the SM particles are localized on the IR brane, the KK graviton couplings and KK graviton masses are given by
\bea
c_{{\rm DM(SM)},n}&=& \left\{ \begin{array}{cc} 1, \qquad {\rm RS}, \vspace{0.3cm} \\  (k_{\rm CW}R) \cdot \frac{n}{m_n R}, \quad {\rm CW},  \end{array} \right. \\
m_n&=& \left\{ \begin{array}{cc} \frac{x_n}{x_1}\,m_G,\quad\qquad {\rm RS}, \vspace{0.3cm} \\   \sqrt{m^2_G+\frac{n^2}{R^2}} , \quad {\rm CW}.  \end{array} \right.
\eea
Here, for RS model, $m_G =x_1\, k_{\rm RS}\, e^{-k_{\rm RS} \pi R}$ with $k_{\rm RS}$ being the AdS curvature scale, and $x_n$ are the zeros of $J_1(x_n)=0$, i.e. $x_n=3.83, 7.02, 10.17, 13.32$ for $n=1,2,3,4$,  which can be approximated to $x_n=(n+1/4)+{\cal O}(n^{-1})$ for $n\gg 1$, and $R$ is the radius of the warped extra dimension. For CW model, $m_G= k_{\rm CW}$ with $k_{\rm CW}$ being the 5D curvature scale. Moreover, the overall suppression scale for massive graviton couplings is
\bea
 \Lambda&=& \left\{ \begin{array}{cc} M_P\, e^{-k_{\rm RS}\pi R}=\frac{M^{3/2}_5}{\sqrt{k_{\rm RS}}}\, e^{-k_{\rm RS}\pi R}, \qquad\qquad {\rm RS}, \vspace{0.3cm} \\  
 M_P \sqrt{k_{\rm CW}\pi R}\, e^{-k_{\rm CW}\pi R}=M^{3/2}_5 \sqrt{\pi R}, \quad {\rm CW}.  \end{array} \right.
\eea
where $M_P, M_5$ are the 4D and 5D Planck masses, respectively, and the relations between them were used in the second equality in each line. Therefore, the KK graviton mass and the KK graviton coupling can be chosen independently, attributed to the choice of the 5D curvature scale ($k_{\rm RS}$ or $k_{\rm CW}$) and the radius of the extra dimension $R$. We note that the ratio of the first KK graviton mass to the suppression scale are given by $\frac{m_G}{\Lambda}=x_1\,\frac{k_{\rm RS}}{M_P}$ in RS model and $\frac{m_G}{\Lambda}=\frac{k_{\rm CW}}{M_5}\,\frac{1}{\sqrt{M_5\pi R}}$ in clockwork model, so the ratio is limited to $\frac{m_G}{\Lambda}\lesssim {\cal O}(1)$ for $k_{\rm RS}\lesssim M_P$ and $k_{\rm CW}\lesssim M_5$, respectively. 

The model dependence of the widths of heavier KK gravitons is discussed in appendix B.
The effects of KK modes of graviton on dark matter physics were discussed in the context of the RS model \cite{GMDM} and the continuum clockwork model \cite{G2}. The impacts of the double and triple interactions of KK gravitons have been discussed in Ref.~\cite{G2} and \cite{clockwork} for dark matter annihilations and decays of heavy KK modes, respectively.
It would be also interesting to generalize the above discussion to the case with more general warped geometries \cite{general-CW}.

In the following, we focus on the minimal interactions of the KK gravitons at the linear level, motivated by the warped extra dimension, and study the quantitative effects of such KK modes on dark matter annihilations into the SM particles and DM elastic scattering processes. 

\subsection{Dark matter annihilations}

First, the KK modes contribute to the $s$-channels of dark matter annihilating into the SM particles by
\bea
(\sigma v)_{{\rm DM\,  DM}\rightarrow {\rm SM\, SM}}= A_s  |S|^2
\eea
where   
\bea
S=\frac{1}{\Lambda^2}\sum_{n=1}^\infty\frac{c_{{\rm DM},n} c_{{\rm SM},n}}{s-m^2_n + i\Gamma_n m_n}\simeq \frac{1}{\Lambda^2}\sum_{n=1}^\infty \frac{c_{{\rm DM},n} c_{{\rm SM},n}}{4m^2_{\rm DM}-m^2_n + i\Gamma_n m_n}
\eea
where $A_s$ is the resonance-independent factors in the cross section. Then, using eqs.~(\ref{s-RS}) and (\ref{s-CW}) in appendix C, we get the modified $s$-channel cross sections of scalar dark matter annihilating into a pair of the SM fermions, whose masses are ignored, as follows:
\bea
(\sigma v)_{SS\rightarrow\psi{\bar\psi} } 
\simeq  v^4 \cdot  \frac{ N_c c^2_S c^2_\psi m_S^6  }{360\pi \Lambda^4 m^4_G} \cdot f(m_S,m_G)
\eea 
with
\bea
f(m_S,m_G)=\frac{x^2_1 m^2_G}{16 m^2_S} \bigg(\frac{J_2(2x_1 m_S/m_G)}{J_1(2x_1 m_S/m_G)} \bigg)^2 \\
\eea
for RS model,
or
\bea
f(m_S,m_G)&=& \frac{ m^4_G}{64m^4_S}
\bigg\{ (k_{\rm CW} \pi R) \,\coth(k_{\rm CW}\pi R) \nonumber \\
&&\quad- \pi R\sqrt{m^2_G-4m^2_S} \,\coth\Big(\pi R\sqrt{m^2_G-4m^2_S}\Big) \bigg) \bigg\}^2
\eea
for CW model.
We note that the $s$-channel resonances in RS model appear at the zeros of $J_1(2x_1 m_S/m_G)$, namely, at $m_G=\frac{2x_1}{x_n}\, m_S$, with $J_1(x_n)=0$, whereas the the $s$-channel resonances in CW model appear only at $m_G=2m_S$. 
The other $s$-channel cross sections for dark matter of other spins into the SM particles and the rest $s$-channel cross sections are modified with the same overall factor, $f(m_S,m_G)$. 
 For $m_{\rm DM}\ll m_G$, the annihilation cross section into the SM fermions is enhanced by $f(m_S,m_G)\simeq 3$ in RS model and it is modified by $f(m_S,m_G)\simeq \frac{(k_{\rm CW}\pi R)^2}{16\pi^4}$ in clockwork model, which is 
about $8$ for $\Lambda=3\,{\rm TeV}$. But, when scalar dark matter and the first KK graviton have similar masses, the contributions from higher KK modes are not significant. 
Similar conclusions can be drawn also for fermion and vector dark matter.

\subsection{Dark matter scatterings}

The contributions of KK gravitons to the $t$-channels of DM-nucleon scattering and DM self-scattering cross sections are given, respectively, by
\bea
\sigma_{{\rm DM\,  SM}\rightarrow {\rm DM\, SM}}= A_t |T_1|^2, \\
\sigma_{{\rm DM\,  DM}\rightarrow {\rm DM\, DM}}= B_t |T_2|^2
\eea
with 
\bea
T_1&=&\frac{1}{\Lambda^2}\sum_{n=1}^\infty\frac{c_{{\rm DM},n} c_{{\rm SM},n}}{t-m^2_n + i\Gamma_n m_n}
\simeq -\frac{1}{\Lambda^2}\sum_{n=1}^\infty\frac{c_{{\rm DM},n} c_{{\rm SM},n}}{m^2_n}, \\
T_2&=&\frac{1}{\Lambda^2}\sum_{n=1}^\infty\frac{c^2_{{\rm DM},n} }{t-m^2_n + i\Gamma_n m_n}
\simeq -\frac{1}{\Lambda^2}\sum_{n=1}^\infty\frac{c^2_{{\rm DM},n}}{m^2_n}
\eea
where $A_t$ is the factor independent of the KK graviton propagator in the cross section, and SM stands for nucleon for WIMP dark matter or electron for light dark matter. 
Similarly, the KK modes contribute similarly to the $t$-channels of DM-electron scattering for direct detection and kinetic equilibrium, with a similar approximate KK graviton propagator for small momentum transfer.
We note that in the case of DM self-scattering, the $t$-channel contributions are dominant in the Born limit, so the above discussion on the $t$-channels would be sufficient.

First, for the DM-nucleus scattering in direct detection, using eqs.~(\ref{t-RS}) and (\ref{t-CW}), we only have to replace the effective nucleon couplings in eqs.~(\ref{fp}) and (\ref{fn}) by the sum of KK modes, as follows,
\bea
f^{\rm DM}_{p,n} =\frac{c^{p,n}_{\rm eff} c_{\rm DM} m_N m_{\rm DM}}{4 \Lambda^2}\times \left\{ \begin{array}{cc} \sum_n \frac{1}{m^2_n}=\frac{x^2_1}{8m^2_G} ,  \quad {\rm RS}, \vspace{0.3cm} \\   
(k_{\rm CW} R)^2\sum_n \frac{n^2}{m^{4}_n R^2}\approx  \frac{\pi (k_{\rm CW}R)^3}{4 m^2_G},\quad {\rm CW}. \end{array} \right.
\eea
Second, for the DM-electron scattering in direct detection, we can similarly replace the corresponding cross section in eq.~(\ref{DDe}) by
\bea
\sigma_{\rm DM-e} ={4 c_e^2 c_{\rm DM}^2 m_e^4 \over 9\pi \Lambda^4 (m_e+m_{\rm DM})^2}
 \times \left\{ \begin{array}{cc} \Big(\sum_n \frac{1}{m^2_n} \Big)^2=\frac{x^4_1}{64 m^4_G},  \quad {\rm RS}, \vspace{0.3cm} \\   
\Big((k_{\rm CW} R)^2\sum_n \frac{n^2}{m^{4}_n R^2}\Big)^2\approx  \frac{\pi^2 (k_{\rm CW} R)^6}{16 m^4_G},\quad {\rm CW}. \end{array} \right.
\eea
Moreover, the momentum relaxation rate for kinetic equilibrium  in eq.~(\ref{relax}) becomes 
\bea
\gamma_{{\rm DM}\, e\rightarrow {\rm DM} \,e}={127 \pi^5 c_e^2 c_{\rm DM}^2 m_{\rm DM}  \over 270 \Lambda^4}\, T^8 \times \left\{ \begin{array}{cc} \Big(\sum_n \frac{1}{m^2_n} \Big)^2=\frac{x^4_1}{64 m^4_G},  \quad {\rm RS}, \vspace{0.3cm} \\   
\Big((k_{\rm CW} R)^2\sum_n \frac{n^2}{m^{4}_n R^2}\Big)^2\approx  \frac{\pi^2 (k_{\rm CW} R)^6}{16 m^4_G},\quad {\rm CW}. \end{array} \right.
\eea

Finally, for the DM self-scattering, the corresponding $t$-channel cross sections in the Born limit are also modified due to the KK modes, as follows,
\bea
\sigma_{\rm DM\, self}= \frac{c^4_{\rm DM} m^6_{\rm DM}}{18\pi \Lambda^4}\, \times \left\{ \begin{array}{cc} \Big(\sum_n \frac{1}{m^2_n} \Big)^2=\frac{x^4_1}{64 m^4_G},  \quad {\rm RS}, \vspace{0.3cm} \\   
\Big((k_{\rm CW} R)^2\sum_n \frac{n^2}{m^{4}_n R^2}\Big)^2\approx  \frac{\pi^2 (k_{\rm CW} R)^6}{16 m^4_G},\quad {\rm CW}. \end{array} \right.
\eea

As a consequence, for RS model, the contributions of the heavier KK modes to the $t$-channel scattering cross sections for dark matter are about 3.4 larger than the one of the first KK mode only. 
For CW model, the contributions from the heavier KK modes depend on the warp factor, that is, they can be important for  $k_{\rm CW}R\gtrsim 1.1$, independent of the spins of dark matter. 
Therefore, in both models, we can make the direct detection bounds less stringent on the couplings of the first KK graviton by including the heavier KK modes for the $t$-channel scattering processes.

\subsection{KK graviton productions}

Each of heavier KK modes of graviton can be also singly produced with a sufficiently large center of mass energy at LHC, with similar signatures as for the first KK graviton. 
However, in clockwork model, the KK graviton masses can be almost degenerate, namely, the mass difference between the $n+1$-th and $n$-the KK graviton masses is given by $\Delta m_n\equiv m_{n+1}-m_n= m_G (2n+1)/(2(kR)^2)\ll m_G$ for $kR\gg 1$. In this case,  almost continuum KK gravitons can be produced simultaneously, leading to the photon or lepton energy spectrum of periodic shape \cite{clockwork-pheno,CMS-clockwork}.

As we discussed in Section 4.4, another smoking-gun signal for the spin-2 mediator would be through $e^+e^-\rightarrow \gamma\, G$ or $q{\bar q}\rightarrow g\, G$, which could identify the signatures of spin-2 mediator couplings. 
For $s\gg m^2_G$,  the heavier KK modes can be also produced at the LHC. 
In RS model, the KK graviton masses are well separated, so we could search for the heavier KK modes as for the first KK graviton as we discussed in Section 4. On the other hand, in clockwork model,  almost continuum KK gravitons could be produced against mono-jet, decaying visibly or invisibly, so the resulting experimental signatures could be significantly different from those in the effective theory only with a single spin-2 mediator case.

\subsection{Non-linear interactions of spin-2 mediator}

As we mentioned in the beginning of the section, there also appear non-linear interactions of KK gravitons in the 4D effective theory, contributing to the DM annihilation channels, such as ${\rm DM\,DM}\rightarrow GG$. 
There have been attempts to tackle the unitarity bound on the non-linear interactions of a massive spin-2 particle in the dRGT realization of the massive spin-2 particle \cite{dRGT,adam} or include the non-linear interactions in the scattering amplitudes of KK gravitons in the RS model \cite{RS-GG}.

In this section,  we discuss briefly the effects of non-linear interactions on the unitarity bound from ${\rm DM\,DM}\rightarrow GG$ or ${\rm DM}\,G\rightarrow {\rm DM}\,G$ by crossing symmetry, in a model-independent way of realizing the massive spin-2 particle.

The perturbative unitarity can give an important constraint on the effective theory for the massive spin-2 particle. 
In particular, for the dark matter annihilation into a pair of spin-2 mediators, the unitarity scale depends on other couplings of the spin-2 mediators such as quadratic couplings to dark matter and cubic self-couplings \cite{dRGT,adam}. In particular, non-linear interactions for the massive spin-2 particle are important for the ghost-free realization of a massive spin-2 particle \cite{dRGT,dRGT-review}.

Fixing the quadratic coupling to dark matter and cubic self-couplings for the massive spin-2 mediator appropriately in the dRGT gravity \cite{dRGT}, the unitarity for ${\rm DM}\,G\rightarrow {\rm DM}\,G$ or ${\rm DM\,DM}\rightarrow GG$ by crossing symmetry can be preserved best until the energy scale \cite{adam}, given by
\bea
E_{\rm max}\sim \bigg(\frac{m_G\Lambda^2}{c^2_{\rm DM}}\bigg)^{1/3}.  \label{unitarity}
\eea
This result is in contrast with the case without non-linear interactions for which unitarity would be violated at  $E'_{\rm max}\sim (m^2_G\Lambda/c_{\rm DM})^{1/3}$ \cite{adam}, which is parametrically smaller that the one in the dRGT gravity for a light spin-2 mediator. 
Therefore, in the dRGT realization of the ghost-free spin-2 mediator, we require $E_{\rm max} \gtrsim m_{\rm DM}$ at least in the regime where the DM annihilation processes are relevant, in other words,
\bea
\frac{\Lambda}{c_{\rm DM}} \gtrsim \bigg(\frac{m^3_{\rm DM}}{m_G}\bigg)^{\frac{1}{2}}.
\eea
As a consequence, we have checked that the above unitarity constraint is satisfied in most of the parameter space for dark matter in the previous sections.
It would be interesting to perform the detailed calculations of ${\rm DM\,DM}\rightarrow GG$ in the dRGT effective theory of the massive spin-2 mediator with non-linear interactions or in the specific benchmark models with the warped extra dimension that we considered in this section, but we plan to revisit this important issue in a future work.

\section{Conclusions}

We have explored the general production mechanisms for WIMP and sub-GeV scale light dark matter with arbitrary spin in the scenarios of gravity-mediated dark matter.  The spin-2 mediator interactions of dark matter as well as SM particles are constrained by direct and direct detections, precision measurements and collider experiments. We showed that the parameter space where dark matter annihilates dominantly into the SM fermions is disfavored, due to direct detection and LHC dijet bounds for weak-scale WIMP case, and mono-photon searchers at BaBar experiments for light dark matter. On the other hand, we found that when dark matter annihilates dominantly into a pair of spin-2 particles in both allowed and forbidden regimes, the model is consistent with current bounds from direct detection and collider experiments. In particular, light dark matter with forbidden channels is not constrained by current indirect detection and CMB measurements. 

As compared to the papers on this topic in the literature, the new ingredients of this article are summarized. We made a complete analysis of the DM-nucleon elastic scattering by taking into account gluon couplings at tree level and loop corrections from heavy quarks and thus extend the previous results in Ref.~\cite{GMDM-dd} significantly. We also provided the new results for forbidden and $3\rightarrow 2$ annihilation channels for light dark matter, DM self-scattering, DM-electron elastic scattering as well as the spin-2 mediator production at linear colliders. 
The new results for the complete treatment of the DM-nucleon elastic scattering is important for constraining WIMP dark matter by XENON1T. On the other hand, the new results for light dark matter are crucial for finding viable models with a light massive spin-2 mediator. In particular, the new forbidden channels make light dark matter compatible with CMB at recombination while the spin-2 mediator has sizable couplings to the SM.
Moreover, we also presented concrete benchmark models for specific masses and couplings for the spin-2 mediators from the warped extra dimension.

\appendix
\section{DM-nucleon scattering amplitudes}

\section*{\large Scalar dark matter}

From the results, we obtain the scattering amplitude between fermion dark matter and nucleon as follows,
\bea
{\cal M}_S &=& \frac{ic_S }{2m^2_G \Lambda^2} \, \bigg\{ 2{\tilde T}^S_{\mu\nu}\,\cdot \frac{1}{m_N} \, \Big(p_\mu p_\nu -\frac{1}{4} m^2_N g_{\mu\nu}\Big)\Big[c_q(q(2)+{\bar q}(2))+c_g G(2)\Big] 
\nonumber \\ 
&&+\frac{1}{6} m_N \Big[c_q \Big(f^N_{Tq}-\frac{2}{27} f_{TG}\Big)+\frac{11}{9} c_g f_{TG} \Big] {T}^S \bigg\} {\bar u}_N(p) u_N(p) \nonumber \\
&=&\frac{ic_S}{2m^2_G \Lambda^2} \, \bigg[ \frac{2}{m_N}\,\Big[c_q(q(2)+{\bar q}(2))+c_g G(2)\Big]  \Big(\frac{1}{2}m^2_N (k_1\cdot k_2)-2(p\cdot k_1)(p\cdot k_2) \Big) \nonumber \\
&&-\frac{1}{3} m_N \Big[c_q \Big(f^N_{Tq}-\frac{2}{27} f_{TG}\Big)+\frac{11}{9} c_g f_{TG} \Big](2m^2_S - k_1\cdot k_2 ) \bigg]{\bar u}_N(p) u_N(p), 
\eea

\section*{\large Fermion dark matter}

The scattering amplitude between fermion dark matter and nucleon can be obtained similarly, as follows,
\bea
{\cal M}_\chi &=& \frac{ic_\chi }{2m^2_G \Lambda^2} \, \bigg\{ 2{\tilde T}^\chi_{\mu\nu}\,\cdot \frac{1}{m_N} \, \Big(p_\mu p_\nu -\frac{1}{4} m^2_N g_{\mu\nu}\Big)\Big[c_q(q(2)+{\bar q}(2))+c_g G(2)\Big] 
\nonumber \\ 
&&+\frac{1}{6} m_N \Big[c_q \Big(f^N_{Tq}-\frac{2}{27} f_{TG}\Big)+\frac{11}{9} c_g f_{TG} \Big] {T}^\chi \bigg\} {\bar u}_N(p) u_N(p) \nonumber \\
&=&  \frac{ic_\chi c_\psi}{2m^2_G \Lambda^2} \,  \bigg\{- \frac{1}{m_N}\,\Big[c_q(q(2)+{\bar q}(2))+c_g G(2)\Big] [p\cdot (k_1+k_2)] ({\bar u}_\chi(k_2) \slashed{p} u_\chi(k_1)) \nonumber \\
&& +\frac{1}{2} m_N m_\chi \Big[ c_q(q(2)+{\bar q}(2))+c_g G(2) \nonumber \\
&&-\frac{1}{3}\Big(c_q \Big(f^N_{Tq}-\frac{2}{27} f_{TG}\Big)+\frac{11}{9} c_g f_{TG}  \Big) \Big] ({\bar u}_\chi(k_2) u_\chi(k_1))\bigg\}{\bar u}_N(p) u_N(p). 
 \eea

\section*{\large Vector dark matter}

The scattering amplitude between vector dark matter and nucleon is also given by
\bea
{\cal M}_X &=&  \frac{ic_X }{2m^2_G \Lambda^2} \, \bigg\{ 2{\tilde T}^X_{\mu\nu}\,\cdot \frac{1}{m_N} \, \Big(p_\mu p_\nu -\frac{1}{4} m^2_N g_{\mu\nu}\Big)\Big[c_q(q(2)+{\bar q}(2))+c_g G(2)\Big] 
\nonumber \\ 
&&+\frac{1}{6} m_N \Big[c_q \Big(f^N_{Tq}-\frac{2}{27} f_{TG}\Big)+\frac{11}{9} c_g f_{TG} \Big] {T}^X \bigg\} {\bar u}_N(p) u_N(p) \nonumber \\&=& \frac{ic_X c_q}{2m^2_G \Lambda^2} \,\epsilon^\alpha(k_1) \epsilon^{*\beta}(k_2) \nonumber \\
&&\quad\times \bigg\{ \frac{2}{m_N}\bigg[ 2p_\alpha p_\beta (k_1\cdot k_2-m^2_X) -\frac{1}{2} m^2_N\eta_{\alpha\beta} (2k_1\cdot k_2-m^2_X) + 2\eta_{\alpha\beta}(p\cdot k_1)(p\cdot k_2) \nonumber \\
&&\quad\quad\quad\quad\quad\quad +m^2_N k_{1\beta} k_{2\alpha}-2p_\alpha k_{1\beta} (p\cdot k_2) - 2p_\beta k_{2\alpha} (p\cdot k_1)  \bigg] \Big[c_q(q(2)+{\bar q}(2))+c_g G(2)\Big] \nonumber \\
&&\quad\quad \quad + \frac{1}{3} m_N m^2_X\Big(c_q \Big(f^N_{Tq}-\frac{2}{27} f_{TG}\Big)+\frac{11}{9} c_g f_{TG}  \Big)   \eta_{\alpha\beta} \bigg\}{\bar u}_N(p) u_N(p).
\eea

\section{Decay widths of spin-2 particles}

The partial decay rates of the KK graviton \cite{GMDM} are given by
\bea
    \Gamma_G(gg)&=& \frac{c_{gg}^2m_G^3}{10\pi \Lambda^2}, \quad\quad
       \Gamma_G(\gamma\gamma)=\frac{c_{\gamma\gamma}^2m_G^3}{80\pi \Lambda^2},  \nonumber \\
      \Gamma_G(ZZ)&=&\frac{m_G^3}{80\pi \Lambda^2}\sqrt{1-4r_Z}\bigg(c_{ZZ}^2+\frac{c_H^2}{12}+\frac{r_Z}{3}\left(3c_H^2-20c_Hc_{ZZ}-9c_{ZZ}^2\right) \nonumber \\
    &+&\frac{2r_Z^2}{3}\left(7c_H^2+10c_Hc_{ZZ}+9c_{ZZ}^2\right) \bigg),   \nonumber  \\\Gamma_G(WW)&=&  \frac{m_G^3}{40\pi \Lambda^2}\sqrt{1-4r_W}\bigg(c_{WW}^2+\frac{c_H^2}{12}+\frac{r_W}{3}\left(3c_H^2-20c_Hc_{WW}-9c_{WW}^2\right) \nonumber \\
    &+&\frac{2r_W^2}{3}\left(7c_H^2+10c_Hc_{WW}+9c_{WW}^2\right)\bigg),   \nonumber \\
    \Gamma_G(Z\gamma)&=& \frac{c_{Z\gamma}^2m_G^3}{40\pi \Lambda^2}(1-r_Z)^3\left(1+\frac{r_Z}{2}+\frac{r_Z^2}{6}\right),   \nonumber 
\eea
\bea
    \Gamma_G(\psi\bar{\psi})&=&\frac{N_c c_\psi^2 m_G^3}{160\pi \Lambda^2} (1-4r_\psi)^{3/2}(1+8r_\psi/3),  \nonumber \\
      \Gamma_G(hh)&=&\frac{c_H^2m_G^3}{960\pi \Lambda^2}(1-4r_h)^{5/2}
\eea
where $c_{\gamma\gamma}=s_{\theta}^2c_2+c_{\theta}^2c_1$, $c_{ZZ}=c_{\theta}^2c_2+s_{\theta}^2c_1$, $c_{Z\gamma}=s_{\theta}c_{\theta}(c_2-c_1)$, $c_{gg}=c_3$, $c_{WW}=2c_2$, $r_i=(m_i/m_G)^2$, and $m_G$ is the lightest KK graviton mass.  

On the other hand, the partial decay rates of the invisible decays of the KK graviton \cite{GMDM} are also given by
\bea
\Gamma(SS)&=&  \frac{(c^G_S)^2 m^3_G}{960 \pi \Lambda^2} \Big(1-\frac{4m^2_S}{m^2_G}\Big)^\frac{5}{2}, \\
\Gamma(\chi{\bar\chi})&=& \frac{ (c^G_\chi)^2 m^3_G}{160 \pi \Lambda^2}
 \left(1-\frac{4m^2_\chi}{m^2_G} \right)^\frac{3}{2} \left(1+\frac{8}{3} \frac{m^2_\chi}{m^2_G}\right), \\
 \Gamma(XX)&=& \frac{ (c^G_X)^2 m^3_G}{960\pi \Lambda^2}\Big(1- \frac{4m^2_X}{m^2_G}\Big)^\frac{1}{2}
\Big(13+\frac{56m^2_X}{m^2_G}+\frac{48m^4_X}{m^4_G}\Big).
\eea

For RS model, the heavier KK modes of graviton couple to the SM particles with the same strength as for the one for the first KK graviton, so we only have to replace the graviton mass by those for the heavier KK modes in the above formulas. Thus, the narrow width approximation holds for the heavier KK modes. 

For CW model, the couplings of the KK modes of graviton are level-dependent, such as $c_{\rm SM(DM),n}=(k_{\rm CW}R)\, n/(m_n R)$ for the SM(DM) particles localized on the IR brane. Thus, the partial decay widths of the KK gravitons scale by the overall factor. For instance, the decay rate of the $n$th KK graviton $G_n$  into a gluon pair becomes
\bea
    \Gamma_{G_n}(gg)=  \frac{n^2 m^2_G}{m^2_n}\, \cdot \frac{c_{gg}^2m_n^3}{10\pi \Lambda^2}=\frac{n^2 m_n}{m_G}\cdot \Gamma_{G_1},
\eea
etc.  The overall factor, $\frac{n^2m_n}{m_G}$, is approximated to $n^2$ for $k_{\rm CW} R\gg 1$,  so the partial decay widths of heavier KK gravitons get enhanced, as compared to the case in RS model with the same coupling for the lightest KK graviton.

\section{The KK sums}

\section*{\large Randall-Sundrum model}

The KK sum relevant for the $s$-channels in RS model is in narrow width approximation
\bea
S&\equiv&  \sum_{n=1}^\infty \frac{1}{m^2_n-s}  \nonumber \\
&=& \frac{M^2_P}{k^2_{\rm RS}\Lambda^2}  \sum_{n=1}^\infty  \frac{1}{x^2_n-sM^2_P/(k^2\Lambda^2)} \nonumber \\
&=&\frac{x_1}{2\sqrt{s}\, m_G}\cdot   \frac{J_2(\sqrt{s} \,x_1/m_G)}{J_1(\sqrt{s}\, x_1/m_G)} \label{s-RS}
\eea
where $J_1(x_n)=0$ and we used 
\bea
\sum_{n=1}^\infty \frac{1}{x^2_n-\sigma^2} = \frac{1}{2\sigma}\cdot \frac{J_2(\sigma)}{J_1(\sigma)}.
\eea

The KK sum relevant for the $t/u$-channels in RS model is given by
\bea
T_1 &\equiv& \sum_{n=1}^\infty\,\frac{1}{m^{2}_n} \nonumber \\
&=& \frac{x^2_1}{m^2_G}\, \sum_{n=1}^\infty\,\frac{1}{x^{2}_n} \nonumber \\
&=&  \frac{x^2_1}{8m^2_G} \label{t-RS}
\eea
where we used $J_n(\sigma)\simeq \frac{1}{n!} \Big(\frac{\sigma}{2} \Big)^n$ for $|\sigma|\ll 1$.
with $m_n= x_n k_{\rm RS}\,e^{-k_{\rm RS}\pi R}$ where $x_n$ are the zeros of $J_1(x_n)$.

\section*{\large Clockwork model}

The KK sum relevant for the $s$-channels in CW model is in narrow width approximation
\bea
S'&\equiv&  \sum_{n=1}^\infty\frac{n^2}{m^2_n R^2}\cdot \frac{1}{m^2_n-s}  \nonumber \\
&=&R^2\sum_{n=1}^\infty \frac{m^2_n R^2-(k_{\rm CW}R)^2}{m^2_n R^2 (m^2_n R^2-s R^2)}
=\frac{R^2}{s}\sum_{n=1}^\infty \bigg(\frac{k^2_{\rm CW}}{n^2+(k_W R)^2}- \frac{k^2_{\rm CW}-s}{n^2+R^2(k^2_{\rm CW}-s)} \bigg)  \nonumber \\
&=& \frac{R^2}{s}  \bigg\{ k^2_{\rm CW} \bigg(\frac{\pi}{2k_{\rm CW}R} \,\coth(k_{\rm CW}\pi R)-\frac{1}{2(k_{\rm CW}R)^2} \bigg) \nonumber \\
&&\quad-\Big(k^2_{\rm CW}-s \Big) \bigg(\frac{\pi}{2R\sqrt{k_{\rm CW}^2-s}} \,\coth\Big(\pi R\sqrt{k_{\rm CW}^2-s}\Big)-\frac{1}{2R^2(k_{\rm CW}^2-s)} \bigg)  \bigg\} \nonumber \\
&=&\frac{1}{2s}\, \bigg\{ (k_{\rm CW}\pi R) \coth(k_{\rm CW}\pi R) -\pi R\sqrt{k^2_{\rm CW}-s}\, \coth\Big(\pi R\sqrt{k_{\rm CW}^2-s}\Big)\bigg\} \label{s-CW}
\eea
where we used $m^2_n R^2= (k_{\rm CW}R)^2+n^2$ and
\bea
 \sum_{n=1}^\infty \frac{1}{n^2+\alpha^2} = \frac{\pi}{2\alpha}\, \coth(\alpha\pi) -\frac{1}{2\alpha^2}.
\eea
For $s>k^2_{\rm CW}$, we only have to replace $\frac{1}{\sqrt{k^2_{\rm CW}-s}}\coth(k_{\rm CW}\pi R)$ by $-\frac{1}{\sqrt{s-k^2_{\rm CW}}}\cot(\pi R \sqrt{k^2_{\rm CW}-s})$. 

For $s\ll k^2_{\rm CW}$, namely, $4m^2_{\rm DM}\ll m^2_G$ for the s-channel annihilations of dark matter, the above KK sum is approximated to
\bea
S'\approx \frac{k_{\rm CW}\pi R}{4k^2_{\rm CW}}\bigg(\coth(k_{\rm CW}\pi R) -\frac{k_{\rm CW}\pi R}{\sinh^2(k_{\rm CW}\pi R)} \bigg).
\eea
Furthermore, for $k_{\rm CW}\pi R\gg 1$, the above result gets more approximated to
\bea
S'\approx  \frac{k_{\rm CW}\pi R}{4k^2_{\rm CW}}=\frac{k_{\rm CW}\pi R}{4m^2_G}.
\eea

The KK sum relevant for the $t/u$-channels in CW model is given by
\bea
T'_1&\equiv &  \sum_{n=1}^\infty\,\frac{n^2}{m^{4}_n R^2} \nonumber \\
&=&  R^2 \sum_{n=1}^\infty\, \frac{n^2}{(n^2+(k_{\rm CW}R)^2)^2} \nonumber \\
&=& \frac{k_{\rm CW}\pi R}{4k^2_{\rm CW}}\,\bigg(\coth(k_{\rm CW}\pi R) -\frac{k_{\rm CW}\pi R}{\sinh^2(k_{\rm CW}\pi R)} \bigg) \label{t-CW}
\eea
where $m_n=\sqrt{k^2_{\rm CW}+n^2/R^2}$, and we used
\bea
 \sum_{n=1}^\infty\,\frac{1}{(n^2+\alpha^2)^2}=\frac{1}{2\alpha^2} \bigg(\frac{\pi}{2\alpha}\,\coth(\alpha\pi)+\frac{\pi^2}{2}\,\frac{1}{\sinh^2(\alpha\pi)}  -\frac{1}{\alpha^2}\bigg).
\eea
For $k_{\rm CW}\pi R\gg 1$, the above sum becomes
\bea
T'_1\approx  \frac{k_{\rm CW}\pi R}{4k^2_{\rm CW}}= \frac{k_{\rm CW}\pi R}{4 m^2_G}.
\eea

\acknowledgments

The work is supported by Basic Science Research Program through the National Research Foundation of Korea (NRF) funded by the Ministry of Education, Science and Technology (NRF-2019R1A2C2003738 and NRF-2018R1A4A1025334). 
The work of YJK is supported in part by the National Research Foundation of Korea (NRF-2019-Global Ph.D. Fellowship Program).


\end{document}